\documentclass[useAMS,usenatbib,aas_macros]{mnras}
\usepackage{graphicx}
\usepackage{amssymb}
\usepackage{footnote}

\usepackage{epsfig}
\usepackage{color}
\usepackage{amssymb}
\usepackage{ams math}
\usepackage[greek,english]{babel}
\usepackage{subfigure}
\usepackage{relsize}
\usepackage{xfrac}
\usepackage{placeins,lscape}
\usepackage{makecell}

\newcommand{\Myr}{\,{\rm Myr}}

\def\hi{\mbox{H\,{\sc i}}}
\def\hii{\mbox{H\,{\sc ii}}}

\def\hbeta{\mbox{H\,{\sc $\beta$}}}
\def\halpha{\mbox{H\,{\sc $\alpha$}}}
\def\oiii{\mbox{[O\,{\sc iii]}}}
\def\oii{\mbox{[O\,{\sc ii]}}}
\def\nii{\mbox{[N\,{\sc ii]}}}

\def\ciii{\mbox{C\,{\sc iii]}}}
\def\ciiiwl{\mbox{C\,{\sc iii]$\lambda\lambda 1907,1909$}}}

\def\lya{\mbox{Ly{\sc $\alpha$}}}
\def\wlya{$W_{\lambda}$(Ly$\alpha$)}

\def\wciii{$W_{\lambda}$(\ciii)}

\newcommand{\zstar}{\,{$Z_{\star}$}}
\newcommand{\zgas}{\,{$Z_{\rm{g}}$}}
\newcommand{\zszo}{\,{$Z_{\star}$/$Z_{\odot}$}}

\newcommand{\mstar}{\,{$M_{\star}$}}
\def\logm{log($M_{\ast}$/M$_{\odot}$)}
\def\logz{log($Z_{\ast}$/Z$_{\odot}$)}

\defcitealias{cullen2019}{C19}
\defcitealias{kornei2010}{K10}

\title[Stellar metallicity as a function of \wlya]{The VANDELS survey: A strong correlation between Ly$\alpha$ equivalent width and stellar metallicity at $\mathbf{3\leq z \leq 5}$}
\author[F. Cullen et al.]{F. Cullen$^{1}$\thanks{E-mail:fc@roe.ac.uk},
R. J. McLure${^{1}}$, 
J. S. Dunlop${^{1}}$,
A. C. Carnall${^{1}}$,
D. J. McLeod${^{1}}$,
\and A. E. Shapley${^{2}}$,
R. Amor\'in${^{3,4}}$,
M. Bolzonella${^{5}}$,
M. Castellano${^{6}}$,
A. Cimatti${^{7,8}}$,
\and M. Cirasuolo${^{9}}$,
O. Cucciati${^{5}}$,
A. Fontana${^{6}}$,
F. Fontanot${^{10}}$,
B. Garilli${^{11}}$,
\and L. Guaita${^{6,12}}$,
M. J. Jarvis${^{13, 14}}$,
L. Pentericci${^{6}}$,
L. Pozzetti${^{5}}$,
M. Talia${^{5,7}}$,
\and G. Zamorani${^{5}}$,
A. Calabr\`o${^{6}}$,
G. Cresci${^{8}}$,
J. P. U. Fynbo${^{15}}$,
N. P. Hathi${^{16}}$,
\and M. Giavalisco${^{17}}$,
A. Koekemoer${^{16}}$,
F. Mannucci${^{8}}$
and A. Saxena${^{6}}$
\\
Affiliations are listed at the end of the paper}

\begin{document}

\date{Accepted -- . Received}

\pagerange{\pageref{firstpage}--\pageref{lastpage}} \pubyear{2020}

\maketitle	

\label{firstpage}

\begin{abstract}
We present the results of a new study investigating the relationship between observed \lya \ equivalent width (\wlya) and the metallicity of the ionizing stellar population (\zstar) for a sample of $768$ star-forming galaxies at $3 \leq z \leq 5$ drawn from the VANDELS survey.
Dividing our sample into quartiles of rest-frame \wlya \ across the range $-58 \rm{\AA} \lesssim$ \wlya \ $\lesssim 110 \rm{\AA}$ we determine \zstar \ from full spectral fitting of composite far-ultraviolet (FUV) spectra and find a clear anti-correlation between \wlya \ and \zstar.
Our results indicate that \zstar \ decreases by a factor $\gtrsim 3$ between the lowest \wlya \ quartile ($\langle$\wlya$\rangle=-18\rm{\AA}$) and the highest \wlya \ quartile ($\langle$\wlya$\rangle=24\rm{\AA}$).
Similarly, galaxies typically defined as Lyman Alpha Emitters (LAEs; \wlya \ $>20\rm{\AA}$) are, on average, metal poor with respect to the non-LAE galaxy population (\wlya \ $\leq20\rm{\AA}$) with \zstar$_{\rm{non-LAE}}\gtrsim 2 \times$ \zstar$_{\rm{LAE}}$.
Finally, based on the best-fitting stellar models, we estimate that the increasing strength of the stellar ionizing spectrum towards lower \zstar \ is responsible for $\simeq 15-25\%$ of the observed variation in \wlya \ across our sample, with the remaining contribution ($\simeq 75-85\%$) being due to a decrease in the \hi/dust covering fractions in low \zstar \ galaxies.
\end{abstract}



\section{Introduction}\label{sec:introduction}

    \begin{figure*}
        \centerline{\includegraphics[width=8in]{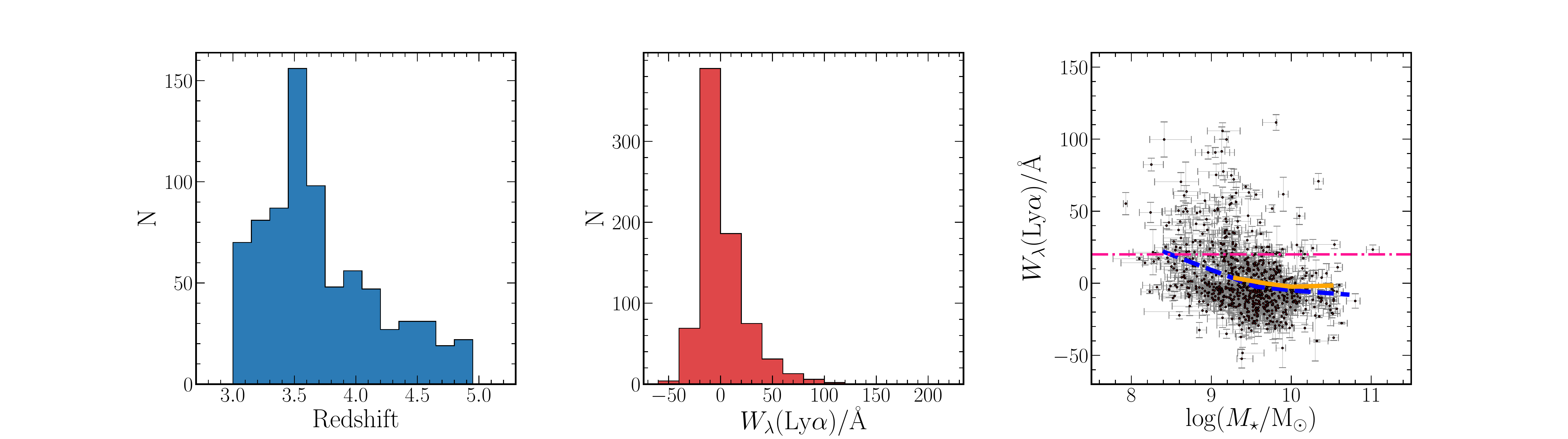}}
        \caption{\emph{Left-hand panel:} Redshift distribution of the $768$ galaxies in our final sample. The redshift range is $3.0 \leq z \leq 5.0$ with a median redshift of $z_{\rm{median}}=3.60$. 
        \emph{Central panel:} The distribution of rest-frame \lya \ equivalent widths; the \wlya \ values fall within the range $-58 \rm{\AA} \lesssim$ \wlya \ $\lesssim 110 \rm{\AA}$ with a median value of \wlya \ $\simeq -4 \rm{\AA}$. 
        \emph{Right-hand panel:} The relationship between galaxy stellar mass and \wlya \ for our sample. 
        The circular data points with error bars are the individual galaxies with the blue dashed curve showing a running average of the data.
        The solid yellow curve shows the $z\simeq3$ \logm $-$ \wlya \ relation derived by \citet{du2018} for comparison.
        The pink dot-dashed line illustrates an equivalent width cut-off commonly used to classify galaxies as LAEs (\wlya $> 20\rm{\AA}$); based on this definition $83\%$ of the LAEs in our sample have \logm $\leq 9.5$, highlighting the fact that \lya \ emission becomes more prevalent at lower stellar masses.}
        \label{fig:sample_properties}
    \end{figure*}

The \lya \ emission line remains a powerful tool for determining the nature of ionizing sources within galaxies as well as probing the ionization state and covering fraction of their interstellar and/or circumgalactic \hi \ gas and dust.
In addition, since \lya \ emission appears to be intrinsically linked to the escape fraction of Lyman continuum (LyC) photons \citep[e.g.][]{nestor2013,verhamme2017,steidel2018,marchi2018}, understanding the nature of \lya \ emitting galaxies will be crucial in characterizing the galaxy population responsible for reionization at $z\gtrsim6$ \citep{fontanot2014,robertson2015}.

The observability of \lya \ depends on the production rate of \lya \ photons within galaxies and on the probability that those photons can escape without being absorbed and/or scattered by the surrounding gas and dust \citep{dijkstra2014}.
Over the past two decades it has become clear that the strength of \lya \ emission increases towards lower mass, relatively dust-free galaxies \citep{kornei2010,hayes2014,cassata2015,hathi2016,oyarzum2017,marchi2019}.
These galaxies typically have low covering fractions of \hi \ gas as inferred from weak Lyman series or low-ionization metal line absorption in their rest-frame UV spectra \citep[e.g.][]{shapley2003,du2018,trainor2019}.
Furthermore, for galaxies with large rest-frame equivalent widths (\wlya), the peak of the \lya \ emission is observed to be closer to a galaxy's systemic redshift, again highlighting the fact that \lya \ emission is favoured when the \hi \ opacity in the immediate vicinity of the galaxy is low \citep{erb2014,trainor2015}. 
In general, these probes of opacity favour higher \lya \ escape fractions in low-mass galaxies.

More recently, signatures of the intrinsic production rate of \lya \ photons have also been explored. 
Using tracers primarily based on rest-frame optical emission-line ratios, various studies have shown that \wlya \ is larger in galaxies containing low-metallicity, high-excitation, \hii \ regions powered by stellar populations that emitting a harder ionizing continuum.
Galaxies selected via line ratios indicative of highly-ionized gas (e.g. \oiii/\oii, \oiii/\hbeta) typically exhibit stronger \lya \ emission than the general population \citep{erb2016,trainor2016,trainor2019}.
Moreover, gas-phase metallicity (\zgas) estimates of \lya \ emitters (LAEs) indicate that they are metal-poor with respect to non-LAEs of similar stellar mass and star-formation rate \citep{charlot1993,finkelstein2011,nakajima2013,song2014,du2019}.
Of course, it is likely that the physical processes governing \lya \ production and escape are fundamentally linked.
For example, we would expect young, low-metallicity, stellar populations that emit harder ionizing spectra to be more effective at ionizing gas in their immediate vicinity, thereby reducing the covering fraction of \hi \ and aiding the escape of \lya \ photons \citep[e.g.][]{erb2010,heckman2011,law2012,erb2014}.
In other words, those galaxies that produce \lya \ photons most efficiently should also be the ones from which those photons have the highest likelihood of escape.

Based on this picture, we expect to observe a correlation between \wlya \ and the stellar metallicity (\zstar) of the ionizing population.
Although existing measurements of \zgas \ strongly hint at such an association, the correlation between \zgas \ and \wlya \ still suffers from various systematics associated with determining \zgas, which become especially severe in the highly-ionized \hii \ regions typical of $z\gtrsim2$ \lya \ emitters, where locally-calibrated line ratio diagnostics may not be applicable \citep{kewley2013, cullen2014,steidel2014,kewley2019}. 
In principle, a cleaner estimate of the metallicity dependence can be made with \zstar, although to date this has never been demonstrated explicitly.
In a recent paper we showed how \zstar \ of the young, hot, O- and B-type stellar populations in high-redshift galaxies can be estimated from rest-frame far-ultraviolet (FUV) spectra using a sample of star-forming galaxies at $2.5 < z < 5.0$ from the VANDELS survey \citep{cullen2019}.
Using this approach, it should be possible to investigate the connection between \wlya \ and \zstar \ and provide a new and powerful constraint on the nature of \lya \ emitting galaxies. 

In this paper, we focus on a subset of VANDELS galaxies with spectral coverage of the \lya \ line to investigate, for the first time, the relationship between \wlya \ and \zstar.
In Section \ref{sec:data} we describe our $z\gtrsim3$ VANDELS star-forming galaxy sample and our method for determining \wlya.
Our determination of the \wlya \ $-$ \zstar \ relation is presented in Section \ref{sec:results} and the implications are discussed in Section \ref{sec:discussion}.
Finally, we list our main conclusions in Section \ref{sec:conclusion}.
Throughout the paper all metallicities are quoted relative to the solar abundance taken from \citet{asplund2009}, which has a bulk composition by mass of $Z_{\ast}=0.0142$,  and all equivalent width values are quoted in the rest-frame with positive values corresponding to emission and negative values to absorption.
We assume the following cosmology: $\Omega_{M} =0.3$, $\Omega_\Lambda =0.7$, $H_0 =70$ km s$^{-1}$ Mpc$^{-1}$.

    \begin{figure}
        \centerline{\includegraphics[width=\columnwidth]{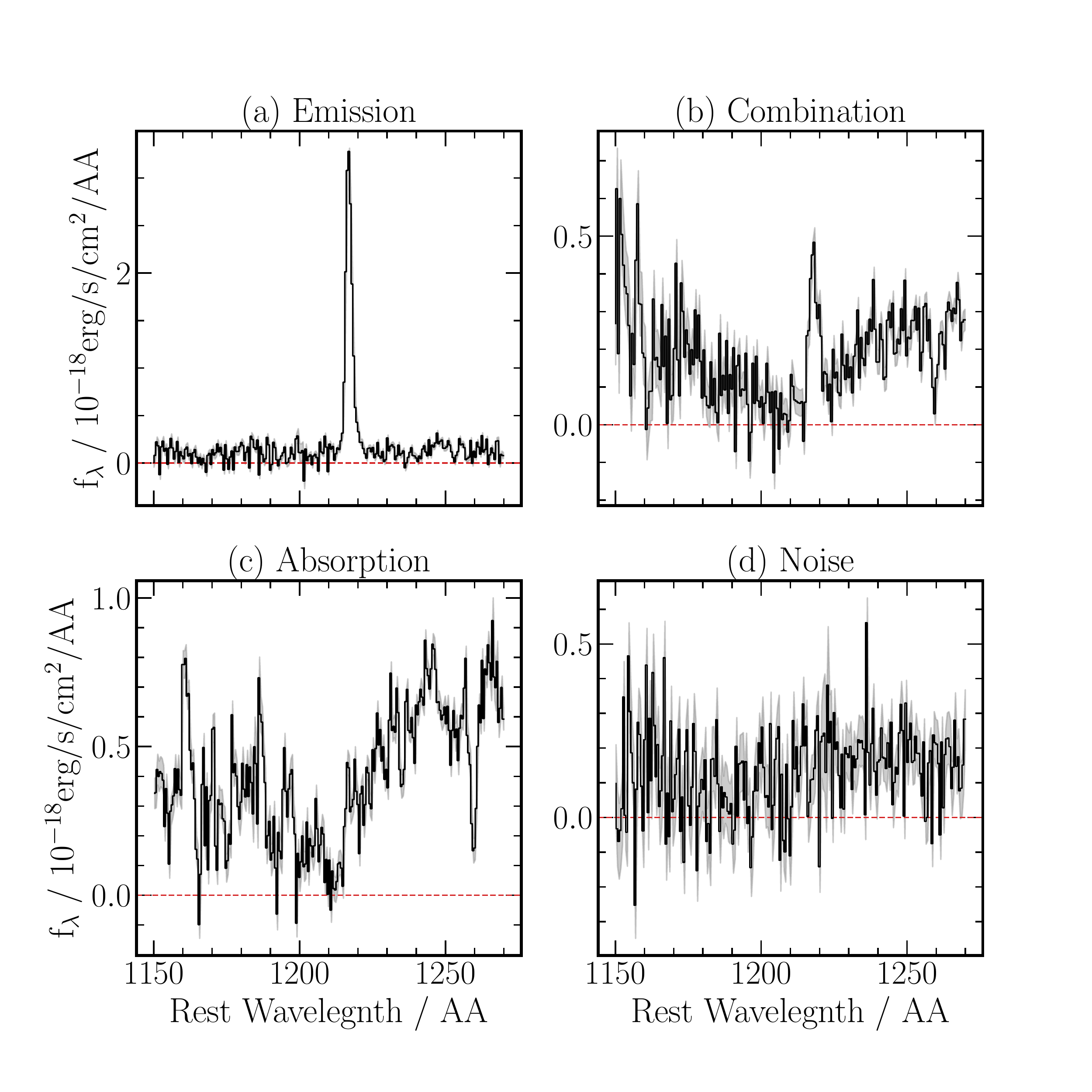}}
        \caption{Example spectra illustrating each of the four \lya \ morphological classifications: emission, combination, absorption and noise.
        For each classification, \wlya \ was calculated following the methods outlined in \citet{kornei2010}.}
        \label{fig:spectal_class}
    \end{figure}

\section{Data and Sample Selection}\label{sec:data}

The spectroscopic data used in this work were obtained as part of the VANDELS ESO public spectroscopic survey \citep{mclure_vandels, pentericci_vandels}.
VANDELS is a deep, optical, spectroscopic survey of the CANDELS \citep{grogin2011,koekemoer2011} CDFS and UDS fields with the ESO-VLT VIMOS spectrograph on ESO's Very Large Telescope (VLT), targeting massive passive galaxies at $1.0 \leq z \leq 2.5$, bright star-forming galaxies at $2.4 \leq z \leq 5.5$ and fainter star-forming galaxies at $3.0 \leq z \leq 7.0$.
VIMOS observations were obtained using the medium-resolution grism which covers the wavelength range $4800 < \lambda_{\rm{obs}} < 10000 \rm{\AA}$ with a resolution of $R=580$ and a dispersion of 2.5 \rm{\AA} per pixel.
The observations and reduction of the VIMOS spectra are described in detail in the first VANDELS data release paper \citep{pentericci_vandels}.

The sample utilized here is drawn from the third VANDELS data release (DR3)\footnote{Data available through the ESO database: http://archive.eso.org/programmatic/$\#$TAP}.
Redshifts for all of the spectra have been determined by members of the VANDELS team and assigned a redshift quality flag ($z_{\rm{flag}}$) as described in \citet{pentericci_vandels}.
In this work, we focus exclusively on star-forming galaxies at $3 \leq z \leq 5$ to ensure both coverage of the \lya \ emission/absorption feature and to enable robust determination of stellar metallicities \citep{cullen2019}. 
All galaxies are required to have a redshift quality flag of $z_{\rm{flag}}=3, 4$ or $9$ (corresponding to a $\geq 95\%$ probability of being correct). 
In total, $777$ galaxies in DR3 satisfy these criteria.
We derived stellar masses for our \lya \ sample using \textsc{fast++}, a rewrite of \textsc{fast} \citep{kriek2009} described in \citet{schreiber2018}.
We employed the \citet{bruzual2003} stellar population synthesis models assuming a \citet{chabrier2003} IMF and delayed exponentially-declining star-formation histories ($M_{\ast} \propto te^{-t/\tau}$) where $t$ is the time since the onset of star formation and $\tau$ is the characteristic star formation timescale.
The age was varied between $7.7 < \mathrm{log}(t/\mathrm{yr}) < 10.1$ in steps of $\Delta(\mathrm{log}(t/\mathrm{yr}))=0.1$, and the star formation timescale was varied between $8.0 < \mathrm{log}(\tau/\mathrm{yr}) < 10.0$ in steps of $\Delta(\mathrm{log}(\tau/\mathrm{yr}))=0.2$.
Dust attenuation was described using the \citet{calzetti2000} attenuation law with $A_{\rm{V}}$ in the range $0.0 \leq A_{\rm{V}} \leq 5.0$ and the stellar population metallicity was allowed to vary between $0.3-2.5 \times \mathrm{Z}_{\odot}$.
The stellar masses of the galaxies in our sample range from $10^8 - 10^{11} \mathrm{M}_{\odot}$ with a median value of $3 \times 10^9 \mathrm{M}_{\odot}$; the median redshift of the sample is $z_{\rm{median}}=3.60$.
The redshift distribution is illustrated in Fig. \ref{fig:sample_properties}.

\subsection{Measuring \lya \ equivalent widths}

To estimate \wlya \ for each galaxy we followed the method described in \citet{kornei2010} \citepalias{kornei2010}.
This technique is designed to provide a robust determination of the wavelength range over which the flux should be integrated based on the morphology of the \lya \ line.
Individual galaxies were visually classified as either `emission', `absorption', `combination' or `noise' (Fig. \ref{fig:spectal_class}; see \citetalias{kornei2010} for a detailed description of the various classifications).
In the first three cases, the peak of the emission/absorption is located and the upper and lower wavelength limits for the flux integration are defined as the wavelength values either side of the peak where the flux intersects an average continuum level.
The blue (i.e. lower) continuum ($c_{\rm{blue}}$) is defined as the median flux value in the range $1120 \rm{\AA} < \lambda_{\rm{rest}} < 1180 \rm{\AA}$ and the red continuum ($c_{\rm{red}}$) as the median value in the range $1228 \rm{\AA} < \lambda_{\rm{rest}} < 1255 \rm{\AA}$.
In the case of `absorption' and `combination' sources, the spectra are first smoothed with a boxcar function of width 6 pixels ($\simeq 3.5 \rm{\AA}$ rest-frame) to minimize the possibility of noise spikes affecting the determination of the upper and lower wavelength boundaries.
For `noise' sources, the \lya \ flux is simply defined as the integrated flux in the range $1200 \rm{\AA} < \lambda_{\rm{rest}} < 1228 \rm{\AA}$.
In all cases the \lya \ line flux is divided by $c_{\rm{red}}$ to yield \wlya.
For $10/777$ objects $c_{\rm{red}}$ was detected at $\leq 2 \sigma$ in the spectrum; in these cases $c_{\rm{red}}$ was estimated using the best fitting SED from the \textsc{fast++} photometric fits.
For each object, this process was repeated 500 times, each time perturbing the flux value in each pixel using its estimated error. 
The value of \wlya \ was taken as the median value of the resulting distribution and the error was estimated using the median absolution deviation (MAD) of the distribution, where $\sigma \simeq 1.4826 \times \mathrm{MAD}$.

    \begin{figure}
        \centerline{\includegraphics[width=\columnwidth]{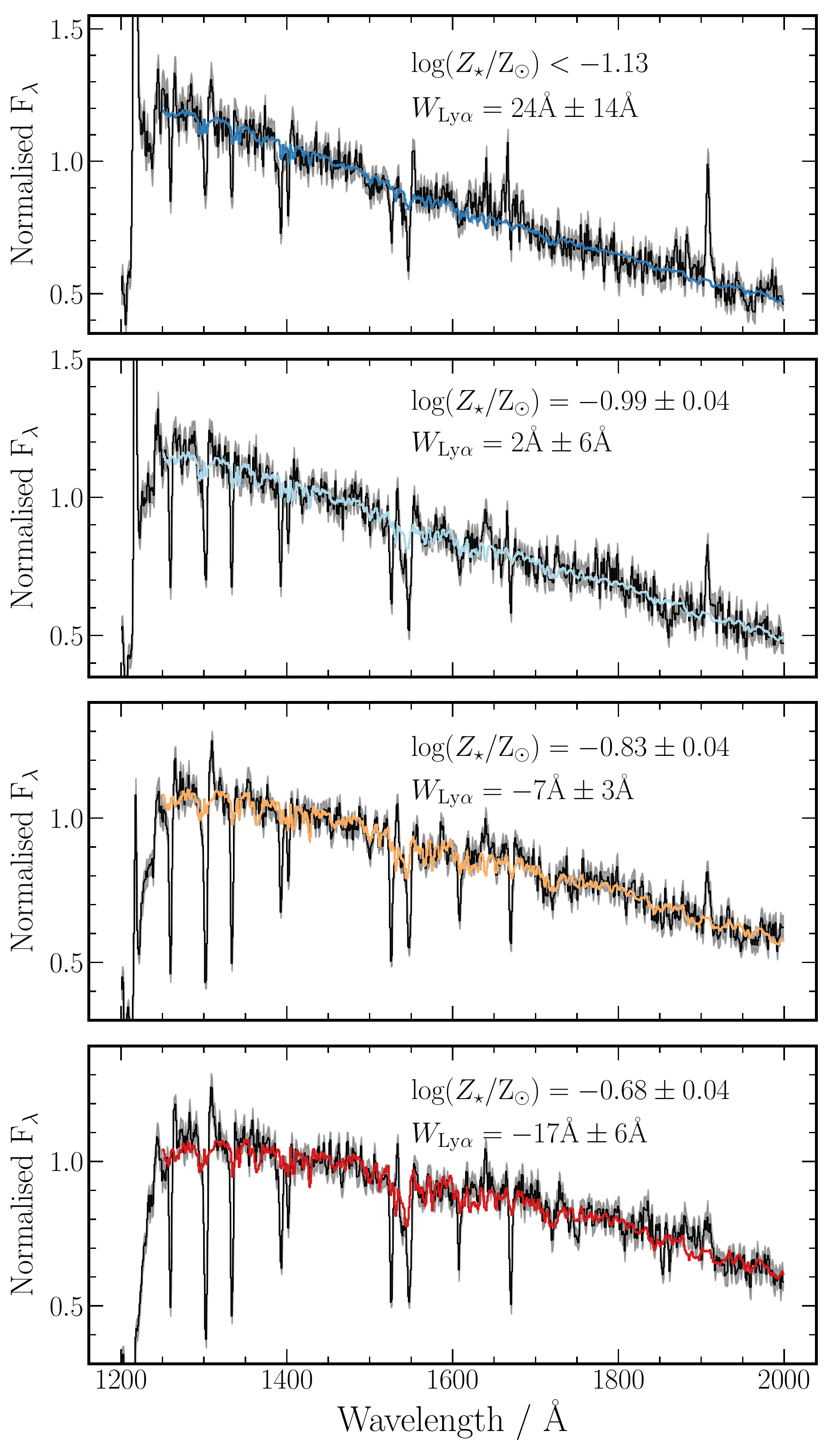}}
        \caption{Composite spectra in four independent \wlya \ quartiles (black lines) with the best-fitting Starburst99 WM-Basic SPS model over-plotted.
        In each panel the median and standard deviation of \wlya \ within the quartile is indicated, along with the best-fitting value of \logz \ and its $1\sigma$ (statistical) error.}
        \label{fig:fuv_fits}
    \end{figure}

    \begin{table*}
    \centering
    \caption{Physical properties and line equivalent widths for the \wlya \ quartiles}\label{table:lya_stack_properties}
    \begin{tabular}{crrcccccc}
        \hline
        \hline
        Quartile & \wlya \ (individual)$^a$ & \wlya (composite)$^b$ & \wciii$^b$ & \logm$^a$ & \logz & $\beta^c$ \\
        \hline
        Q1 & $23.68\pm12.69$ & $19.41\pm0.78$ & $2.89\pm0.23$ &  $9.18\pm0.47$ & $<1.08$ $(68\%)$ & $-1.99\pm0.02$ \\
        Q2 & $2.04\pm6.14$ & $3.76\pm0.22$ & $1.53\pm0.24$ &    $9.45\pm0.47$ & $-0.98\pm0.04$ & $-1.75\pm0.02$ \\
        Q3 & $-7.30\pm3.02$ & $-4.61\pm0.21$ & $1.22\pm0.16$ &   $9.54\pm0.44$ & $-0.82\pm0.04$ & $-1.34\pm0.02$ \\
        Q4 & $-17.69\pm6.12$ & $-18.82\pm0.18$ & $0.84\pm0.37$ & $9.59\pm0.31$ & $-0.69\pm0.04$ & $-1.14\pm0.02$ \\
        \hline
        \multicolumn{7}{l}{$^a$ The median and standard deviation (estimated as $\sigma \simeq 1.4826 \times \mathrm{MAD}$) for the individual galaxies in each quartile.}\\
        \multicolumn{7}{l}{$^b$ Equivalent width values and their associated errors measured directly from the composite spectra.}\\
        \multicolumn{7}{l}{$^c$ UV continuum slope measured directly from the composite spectra following the method described in \citet{cullen2017}.}\\
    \end{tabular}
    \end{table*}

There were $8$ objects for which \wlya \ could not be estimated due to strong contamination at the location of the \lya \ line.
One further object had a extreme equivalent width value of $> 2000 \rm{\AA}$; this object was undetected in the continuum and potentially contaminated by a secondary object in the slit.
These $9$ objects were removed from the sample leaving a total of $768$ galaxies.
Of these, $239/768$ objects were classified as emission spectra, $355/768$ as absorption, $64/768$ as combination and $110/768$ as noise.
The resulting rest-frame equivalent widths span the range $-52 \rm{\AA} \lesssim$ \wlya \ $\lesssim 112 \rm{\AA}$ with a median value of \wlya \ $= -3.6 \rm{\AA}$ as shown in Fig. \ref{fig:sample_properties}.
We note that this median value is slightly lower than the median \wlya \ determined for Lyman Break Galaxy (LBG) selected samples at similar redshifts \citep[c.f. $4\rm{\AA}$,][]{kornei2010} which is a result of LBG selection criteria being biased towards bluer galaxies that exhibit, on average, stronger \lya \ emission (see Section \ref{sec:discussion}).

Fig. \ref{fig:sample_properties} also shows the relation between \wlya \ and stellar mass for the galaxies in our sample.
Consistent with results reported elsewhere in the literature \citep[e.g.][]{oyarzum2016,du2018,marchi2019}, we find a mild anti-correlation such that the average \wlya \ is increasing towards lower stellar-mass galaxies.
We note that this trend should not be a result of observational biases since, by design, the median continuum signal-to-noise ratio (SNR) of the VANDELS spectra is approximately constant as a function of stellar mass \citep{mclure_vandels}. 
Therefore, we are not biased against low-\mstar, low-\wlya \ objects (i.e. objects that would lie in the bottom left-hand corner of the rightmost panel in Fig. \ref{fig:sample_properties}).
In the following section we will return to a discussion of the \wlya-\mstar \ correlation.


\section{Analysis}\label{sec:results}

Armed with the VANDELS spectra, we can explore whether a scaling relation exists between \wlya \ and the stellar metallicity of the young, ionizing, stellar populations in star-forming galaxies at $3.0 \leq z \leq 5.0$.
In this section we first present the \wlya$-$\logz \ relation derived for our sample, followed by a discussion of how this relation compares to the scaling relations observed between both of these quantities and \logm.
Finally, we discuss further insights that can be gained from comparing \wlya \ to the equivalent width of the \ciiiwl \ emission line doublet (\wciii). 

\subsection{\lya \ strength as a function of stellar metallicity}
    \begin{figure}
        \centerline{\includegraphics[width=\columnwidth]{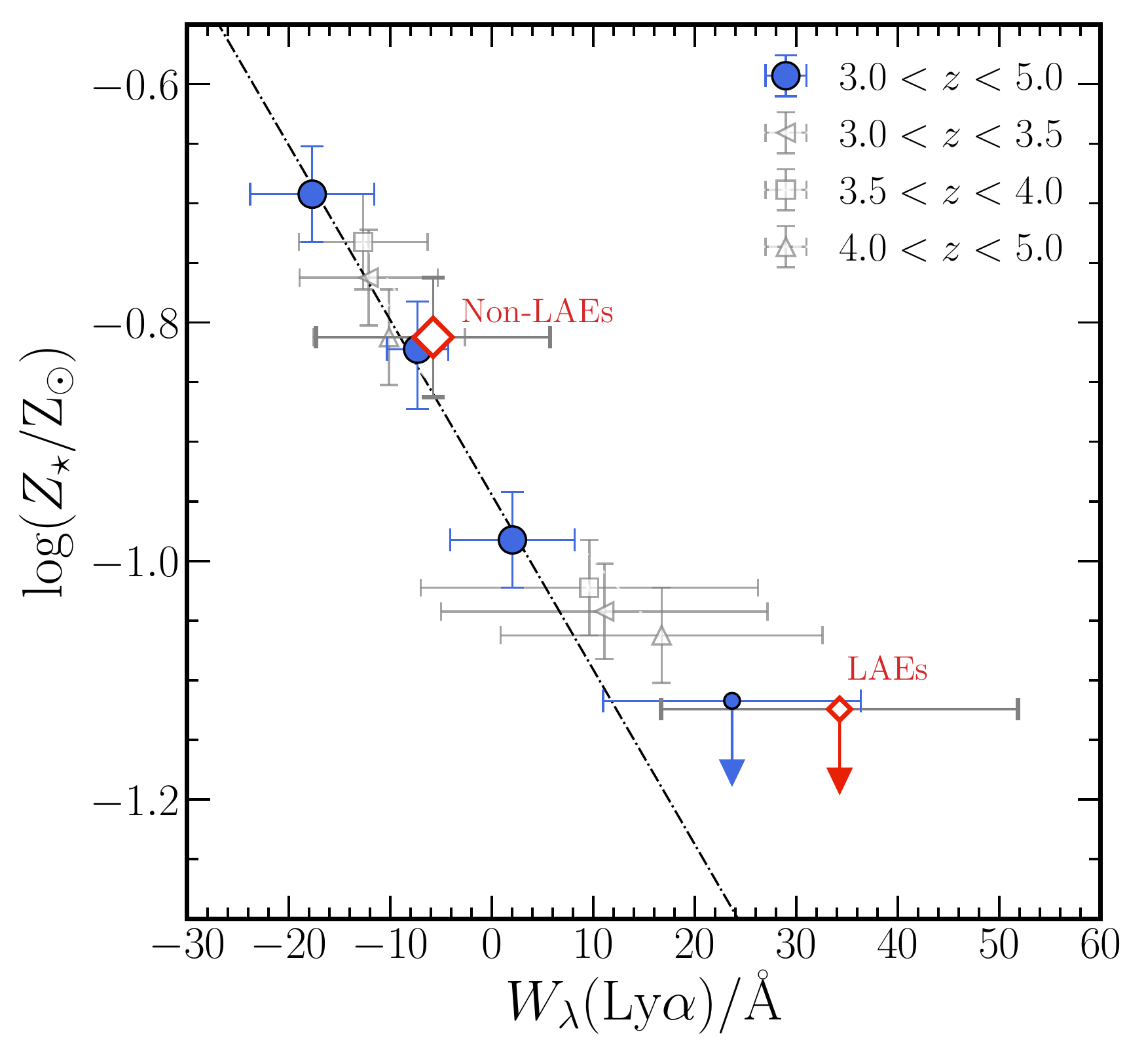}}
        \caption{The relation between \wlya \ and \logz \ for star-forming galaxies at $3 \leq z \leq 5$.
        The blue circular data points with error bars show the data for the full sample split into quartiles of \wlya.
        The \wlya \ values represent the median of all individual \wlya \ values in each quartile.
        The \wlya \ error bars represent the standard deviation of individual \wlya \ values in each quartile (estimated as $\sigma \simeq 1.4826 \times$ median absolute deviation).
        The red diamond data points show the sample split into LAEs (\wlya \ $>20\rm{\AA}$) and non-LAEs (\wlya \ $\leq20\rm{\AA}$).
        Downward pointing arrows represent $68\%$ confidence upper limits on \logz.
        The black dot-dashed line is a log-linear fit to the quartile data excluding the upper limit.
        The open grey data points show the sample split into three redshift bins as indicated in the figure legend (see text for details).}
        \label{fig:logz_vs_lyaew}
    \end{figure}

To assess the dependence of stellar metallicity (\zstar) on \wlya \ we divided our sample into four independent quartiles of \wlya \ and formed composite spectra following the method outlined in \citet{cullen2019} \citepalias{cullen2019}.
Briefly, the individual contributing spectra were first shifted into the rest frame using the measured VANDELS redshift\footnote{Although the VANDELS redshifts are typically measured using \lya \ and/or the ISM absorption lines and therefore do not represent the true systemic redshift of each galaxy, this does not affect the derivation of stellar metallicities from the composite spectra (see \citet{cullen2019} for a discussion).} and then median-combined with an error spectrum estimated via bootstrap re-sampling.
The composite spectra were sampled at $1\rm{\AA}/$pixel and covered the rest-frame wavelength range $1200-2000\rm{\AA}$ with an effective spectral resolution element of $3.0 \rm{\AA}$.
The median and standard deviation (determined from the MAD) of \wlya \ for the four quartiles are given in Table \ref{table:lya_stack_properties}.
The composite spectra are shown in Fig. \ref{fig:fuv_fits} where the transition from net \lya \ emission to net \lya \ absorption can clearly be seen.

Stellar metallicities for the \wlya \ composites were determined following the method described in \citetalias{cullen2019}.
Below we give a brief description of this method, but we refer interested readers to \citetalias{cullen2019} for full details.
We adopted the Starburst99 (SB99) high-resolution WM-Basic stellar population synthesis (SPS) models described in \citet{leitherer2010}, considering constant star-formation models over timescales of 100 \Myr \ with $Z_{\ast}=(0.001, 0.002, 0.008, 0.014, 0.040)$.
To fit the SB99 models to the composite spectra we used a Bayesian nested sampling algorithm implemented in the code \textsc{multinest} \citep{feroz2008,feroz2009}\footnote{We accessed \textsc{multinest} via the python interface \textsc{pymultinest} \citep{buchner2014}.}.
The four parameters in the fit were the stellar metallicity (\zstar) and three dust parameters based on a flexible and physically-motivated form of the attenuation curve described in \citet{salim2018} \citep[see also][]{noll2009}.
The prior in \logz \ was imposed by the SB99 models to be $-1.15 <$ \logz \ $< 1.45$.
Since the models are provided for five fixed metallicity values, we linearly interpolated the logarithmic flux values between the models to generate a model at any metallicity within the prescribed range.
The 1D posterior distribution for \logz \ was obtained by marginalizing over all other parameters in the fit.
The best-fitting \logz \ value was then calculated from the $50$th percentile of this distribution along with the $68\%$ confidence limits.
We note that the errors derived in this way represent the statistical errors for our fitting method and do not account for potential systematic effects related to our choice of SPS model and assumed star-formation history; for a discussion of these issues see \citetalias{cullen2019}.
The best-fitting models for the four \wlya \ stacks are shown in Fig. \ref{fig:fuv_fits} and the best fitting \logz \ values with associated errors are given in Table \ref{table:lya_stack_properties}.

Fig. \ref{fig:logz_vs_lyaew} shows the resulting \wlya \ $-$ \logz \ relation.
The blue circular data points show the four \wlya \ quartiles (Q1-Q4) from Fig. \ref{fig:fuv_fits}, with the downward pointing arrow representing the $68\%$ confidence upper limit on \logz \ for Q1.
We observe a clear correlation between \wlya \ and \logz \ of the form expected: galaxies that exhibit the strongest \lya \ emission contain the lowest metallicity ionizing populations.
Between the lowest and highest \wlya \ quartiles the stellar metallicity decreases from \zszo$=0.20 \pm 0.02$ to \zszo$\lesssim 0.07$ (i.e. greater than a factor 3 at $\simeq 6\sigma$ significance) and the \logz \ $-$ \wlya \ relation (excluding the Q1 upper limit) can be approximately captured by a simple log-liner equation of the form:
\begin{equation}\label{eq:lz_wlya}
\mathrm{log}(Z_{\ast}/\mathrm{Z}_{\odot})=-0.016 (\pm 0.001)W_{\lambda}(\rm{Ly}\alpha)-0.95(\pm0.01).
\end{equation}

As a further check, we also produced composite spectra for the LAEs (\wlya \ $>20\rm{\AA}$) and the non-LAEs (\wlya \ $\leq20\rm{\AA}$) in our sample.
The red open diamonds in Fig. \ref{fig:logz_vs_lyaew} show the average \logz \ and \wlya \ for these populations, which are fully consistent with the quartile data.
For our sample, the ionizing stellar population of non-LAE's is  $\gtrsim 2 \times$ more metal enriched than for the LAE population. 
Again, however, we can only place an upper limit on \logz \ for the LAEs.
In general, the fact that it is only possible to set an upper limit on \zstar \ for the highest \wlya \ galaxies highlights the fact that high-resolution stellar populations at \zszo \ $<< 10\%$ will be required for modeling the low-mass, low-metallicity, galaxy population likely to have played a significant role in \hi \ reionization at $z\gtrsim6$.

We can rule out the possibility that the observed \logz$-$\wlya \ relation is simply a product of differences in the median stellar mass of the \wlya \ quartiles.
This could potentially be an issue because of the known correlation between \zstar \ and \mstar \ \citep[i.e. the stellar MZR][]{gallazzi2005,cullen2019}.
However we find, at least for quartiles Q2-Q4, that the stellar mass distributions have similar median values and variance (Table \ref{table:lya_stack_properties}).
The highest \wlya \ quartile (Q1) has a slightly lower median \mstar \ value, although there is still significant overlap with Q2-Q4 given the large variance within each bin. 
Overall, there is no strong evidence to suggest that the change in \zstar \ with \wlya \ is being driven by differences in the stellar mass distributions of the composite spectra.

   \begin{figure}
        \centerline{\includegraphics[width=\columnwidth]{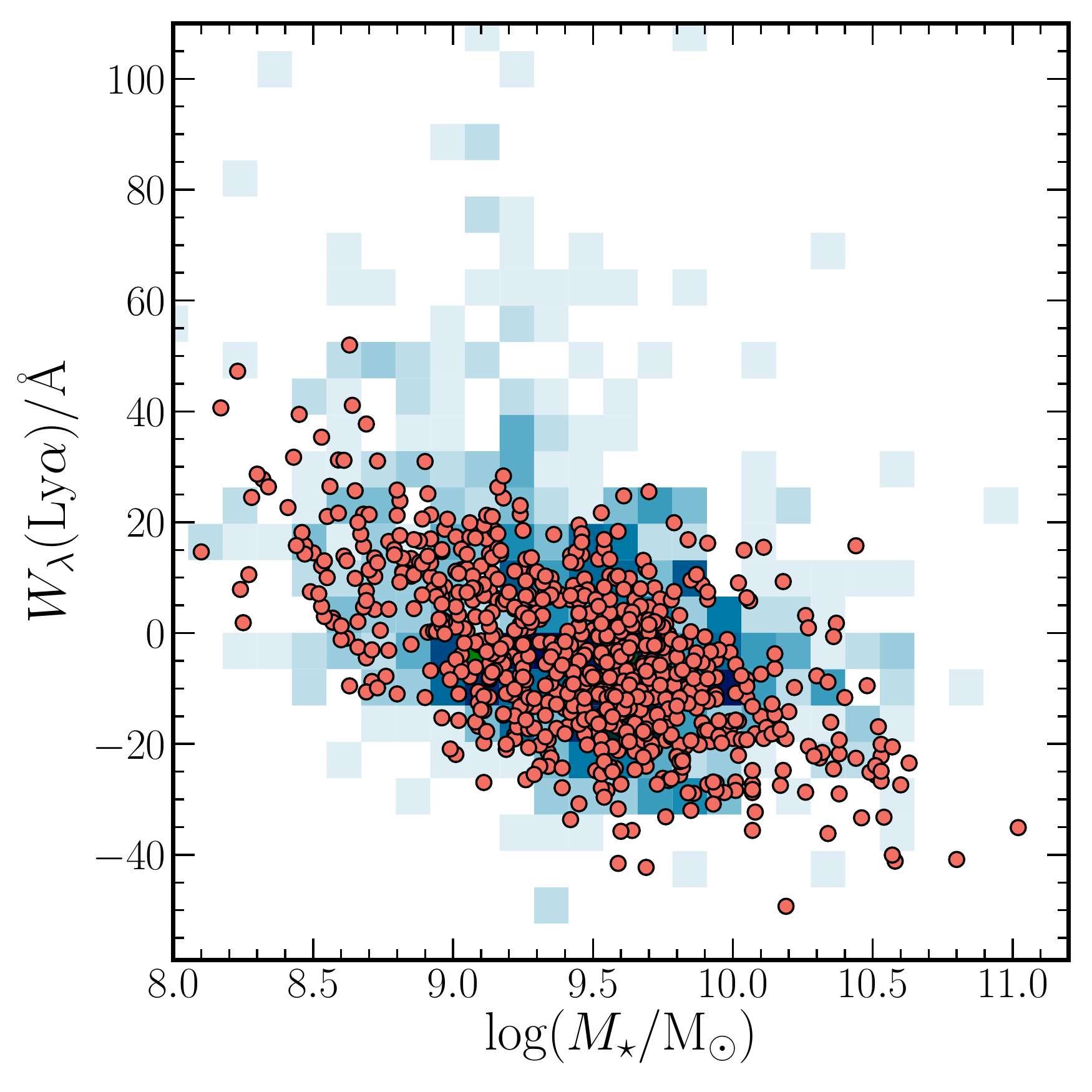}}
        \caption{A comparison between the observed and simulated \logm$-$\wlya \ relations.
        The underlying 2D histogram shows the observed distribution for the VANDELS sample (see also Fig. \ref{fig:sample_properties}) and the orange circular points show a simulated distribution derived using the \logz$-$\wlya \ relation (Fig. \ref{fig:logz_vs_lyaew} and Equation \ref{eq:lz_wlya}) in combination with the stellar mass-metallicity relation from \citet{cullen2019} (see text for details).}
        \label{fig:simulation}
    \end{figure}

Finally, we checked for potential biases introduced by the relatively broad redshift distribution of our sample.
A redshift bias could be a result of (i) a strong dependence of \zstar \ on redshift, or (ii) the increasing IGM attenuation with redshift affecting the relative \wlya \ values across our sample.
As a basic test, we created six stacked spectra across three redshift bins, splitting each redshift bin into two bins of \wlya.
As shown in Fig. \ref{fig:logz_vs_lyaew}, the \wlya$-$\zstar \ relation at each redshift is fully consistent with the relation across the full redshift range.
This strongly suggests no systematic evolution of \zstar \ within our sample, consistent with the results presented in \citet{cullen2019}.
The effect of IGM absorption is more difficult to quantify however, since each galaxy will have its own unique sightline through the IGM.
Nevertheless, it is expected that, on average, the galaxies at higher redshift will have a larger proportion of their \lya \ flux blueward of $1216\rm{\AA}$ attenuated by neutral \hi \ clouds along the line of sight \citep[e.g.][]{pahl2020}.
This could potentially affect how the sample is binned by observed \wlya.
As a simple test we corrected \wlya \ of each galaxy using the relation between \lya \ transmission and redshift reported in \citet{songaila2004}.
Splitting this IGM-corrected \wlya \ distribution into quartiles has a very minor effect on the galaxies assigned to each quartile, and does not change the derived \zstar, although the median \wlya \ values are clearly slightly larger.
Unfortunately, it is not possible to determine the unique IGM correction for each galaxy, and in practice observed \wlya \ is the only measurable quantity.  
Overall, we do not expect any strong redshift biases to be affecting the relation between observed \wlya \ and \zstar.

\subsection{Linking equivalent width, metallicity and mass}

In \citetalias{cullen2019} we presented the relation between \logz \ and \logm \ (i.e. the stellar MZR) for VANDELS star-forming galaxies at $2.5 \leq z \leq 5.0$.
It is interesting to test whether this relation and the \logz$-$\wlya \ relation presented here are consistent with the observed distribution of \logm \ and \wlya \ for the individual galaxies shown in Fig. \ref{fig:sample_properties}.
We note that, although the samples used here and in \citetalias{cullen2019} are not fully independent, the three parameters of interest (\zstar, \mstar, \wlya) have been determined independently, and therefore consistency between the three resulting scaling relations would provide (i) evidence for the robustness of our parameter estimates and (ii) further insight into the nature of \lya \ emission.

To test whether the three relations are self-consistent we performed a simple simulation.
The \citetalias{cullen2019} MZR, which can be approximated by an equation of the form
\begin{equation}\label{eq:mzr}
\mathrm{log}(Z_{\ast}/\mathrm{Z}_{\odot})=0.30(\pm0.06)\mathrm{log}(M_{\ast}/\mathrm{M}_{\odot})+3.7(\pm0.6),
\end{equation}
was used to generate a value of \logz \ for each galaxy in our sample, with an additional scatter of $\sigma_{\rm{log}(Z_{\ast}/\rm{Z}_{\odot})}=0.1$ dex.
Based on the \logz \ value, a value of \wlya \ was generated using Equation \ref{eq:lz_wlya}, again adding a scatter of $\sigma_{W_{\lambda}(Ly\alpha)}=10\rm{\AA}$\footnote{The values of the scatter in \logz \ and \wlya \ were tuned to return a reasonable reproduction of the observed data.}.
The resulting distribution of simulated \wlya$-$\logm \ data is shown overlaid on top of the observed distribution in Fig. \ref{fig:simulation}. 
It can be seen that the bulk of observed \wlya \ values are well-recovered, demonstrating an encouraging consistency between the three independently-measured quantities and highlighting the clear connection between the stellar mass of a galaxy, the metallicity of its young, ionizing, stellar population, and the emergent \lya \ emission.

However, it is interesting to note that this simple model fails to account for the large \wlya \ values ($\gtrsim 50 \rm{\AA}$; $\simeq5\%$ of the full sample) typically seen in galaxies with \logm \ $\lesssim 9.5$.
At these values of \wlya, the MZR and \logz$-$\wlya \ relations would predict significantly lower values of \logm \ than are observed.
This failure of the model could be a result of a number of factors.
Most obviously, the relations provided above are probably not applicable at the lowest stellar mass and \wlya \ values in our sample, where at present we can only estimate upper limits on \zstar.
Placing absolute constraints on \zstar \ in this \logm/\wlya \ regime will likely reveal that a more complex functional form is required to capture the true relations.
Moreover, some of the physical assumptions used in our derivation of \zstar, which is based purely on analysing composite spectra, may not be applicable on a galaxy-by-galaxy basis.
For example, the large \wlya \ values seen in some low mass galaxies may be a result of recent bursts on star formation \citep[e.g.][]{matthee2017} which elevate \wlya \ with respect to the constant star formation histories assumed in our analysis.
However, as this phenomenon only affects a small percentage of our full sample, we defer a more detailed analysis to a future work.
Overall, it is clear that this simple model works remarkably well within the \logm/\wlya \ range for which we can robustly determine \zstar.

Finally, it is interesting to note that the observed distribution can be recovered assuming relatively small values for the scatter in \logz \ and \wlya, implying a perhaps surprisingly small intrinsic scatter for these relations.
Again, this is something we that we will be able to investigate in more detail in a future work utilizing the full VANDELS dataset.

\subsection{The correlation with \ciii \ emission}

Another prominent FUV emission feature, visible in Fig. \ref{fig:fuv_fits}, is the \ciiiwl \ emission line doublet.
Theoretical models predict that the emergent \ciiiwl \ emission will increase towards lower \zstar \ due to the increasing strength and hardness of the ionizing stellar continuum, which regulates both the gas temperature and ionization of $\rm{C}^+$ within \hii \ regions \citep{jaskot2016,senchyna2017,schaerer2018,nakajima2018}.
A variety of previous studies have reported a positive correlation between \wlya \ and \wciii \ \citep[e.g.][]{shapley2003,stark2014,rigby2015,du2018,lefevre2019} and it can clearly be seen from Fig. \ref{fig:fuv_fits} that we observe a similar trend.

To quantify the relation, we measured \wlya \ and \wciii \ directly from the composite spectra.
\wlya \ was measured using the same method as for the individual spectra, and the values with their 1$\sigma$ error bars are reported in Table \ref{table:lya_stack_properties}. 
\wciii \ was measured by first subtracting a local continuum the in the region of the \ciii \ line and measuring the flux from the continuum-subtracted spectra; this flux was then divided by the average absolute continuum value in the wavelength range $1930-1950\rm{\AA}$.
The final value of \wciii \ and its associated $1\sigma$ error bar was calculated using the same Monte Carlo approach adopted for the \lya \ line measurements.
Again, these values are reported in Table \ref{table:lya_stack_properties}.

\begin{figure}
        \centerline{\includegraphics[width=\columnwidth]{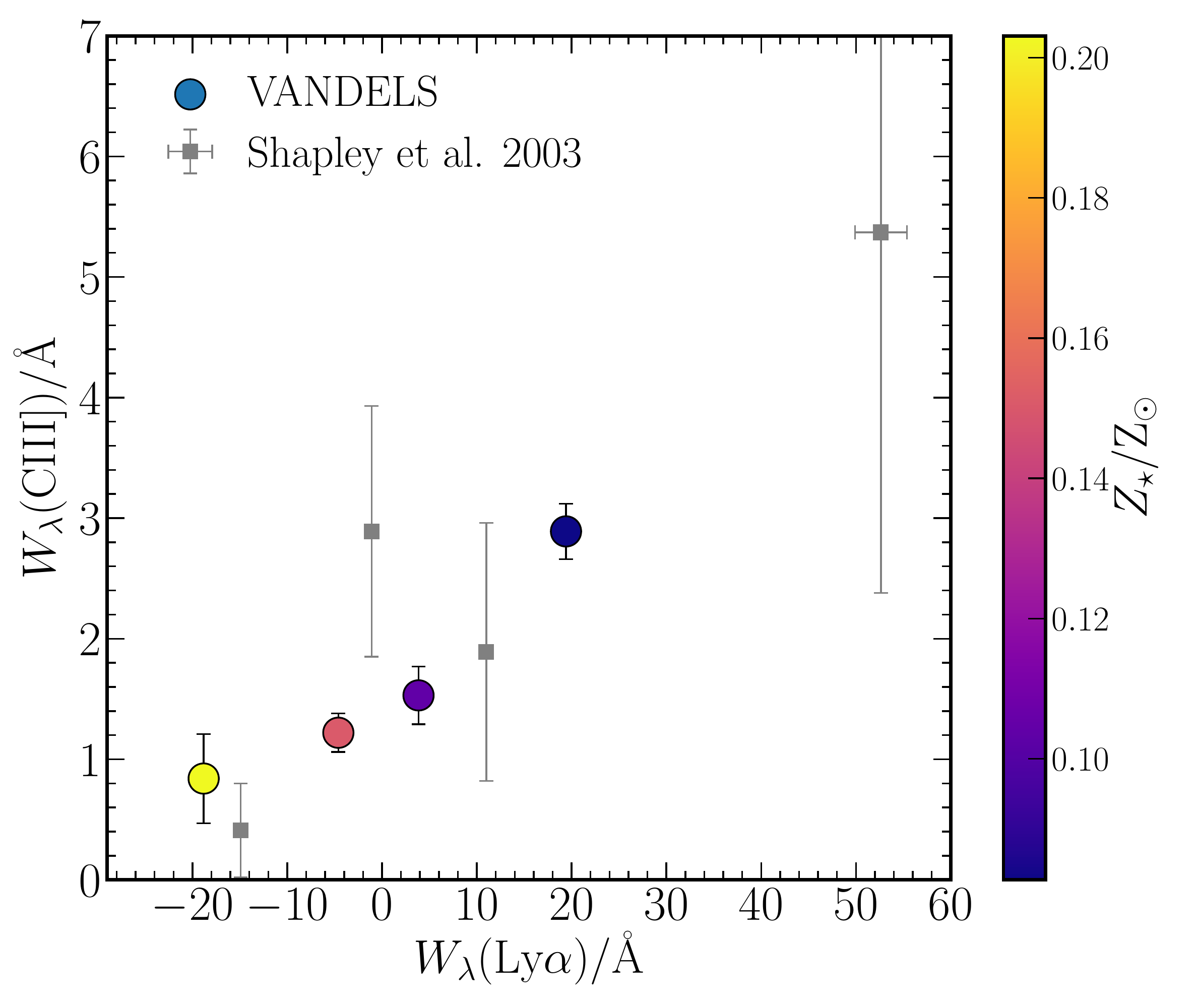}}
        \caption{The relation between \wlya \ and \wciii.
        The circular data points with error bars show the results of our sample split into \wlya \ quartiles colour-coded by the best-fitting stellar metallicity.
        In this case, the values of \wlya, \wciii, and their respective errors are measured directly from the composite spectra as discussed in the text.
        The grey data points are values measured from composite spectra at similar redshifts from \citet{shapley2003}.}
        \label{fig:lya_ciii_relation}
    \end{figure}

The results are shown in Fig. \ref{fig:lya_ciii_relation}, where it can be seen that we find a clear positive correlation between \wlya \ and \wciii. 
This trend is consistent with the results of \citet{shapley2003} and \citet{du2018} at similar redshifts, with \wciii \ increasing by a factor 3 as \wlya \ evolves from $\simeq -20\rm{\AA}$ to $20\rm{\AA}$.
Moreover, as the equivalent width of both lines increase, \zstar \ decreases.
The trend we observe is therefore consistent with a scenario in which the hard ionizing SED of low metallicity stars is closely connected to the observed strength of both the \lya \ and \ciii \ emission lines, which we discuss in more detail below.
We also note that our composite spectra show no evidence for extreme \wciii \ values indicative of AGN photoionization \citep[$\gtrsim10\rm{\AA}$,][]{nakajima2018}.
Finally, it is worth noting that our results also imply that the strength of both \lya \ and \ciii \ emission in galaxies should increase towards higher redshifts as the metallicity of stellar populations decreases further.
Although the visibility of \lya \ will be impeded by an increasing IGM \hi \ fraction at $z > 5$, the \ciii \ line should remain a promising line for study in the reionization era \citep[e.g.][]{stark2014, stark2017}.

\section{Discussion}\label{sec:discussion}

The results presented above have demonstrated, for the first time, a direct correlation between \wlya \ and \zstar \ of the young O- and B-type stellar populations in high-redshift star-forming galaxies.
In this section we briefly discuss this result with respect to other recent investigations of \lya \ emission at high-redshift and finally consider the relative importance of intrinsic production/escape in governing the observed \wlya.

\subsection{Factors governing the observed \wlya}

As discussed in Section \ref{sec:introduction}, the observed \wlya \ is dependent on both the production efficiency of \lya \ photons within galactic \hii \ regions, and on the likelihood that these photons can escape the surrounding ISM/CGM.
In this respect, a strong correlation between \wlya \ and \zstar \ is perhaps unsurprising.
Stellar population synthesis models predict that the ionizing flux of a stellar population increases as stellar metallicity decreases \citep[e.g.][]{schaerer2003,stanway2016}.
An increase in the ionizing flux will naturally lead to an increase in the number of \lya \ photons produced per unit star formation in lower metallicity galaxies.
The increasing strength of the \ciiiwl \ emission line in tandem with \lya \ also supports the idea that the harder ionizing continuum produced by low metallicity stellar populations is crucial in producing large \wlya.
In addition, an increase in the ionizing photon flux may reduce the covering fraction or column density of neutral hydrogen, easing the escape of \lya \ photons \citep[e.g.][]{erb2014}.

This picture is generally supported by previous studies that have correlated \wlya \ with proxies of the ionizing flux and gas-phase metallicity.
Most recently, \citet{trainor2019} have shown that, as well as anti-correlating with the strength of low-ionization UV absorption lines, \wlya \ correlates with the \oiii/\hbeta \ and \oiii/\oii \ nebular emission line ratios in star-forming galaxies at $2\lesssim z \lesssim3$.
Both of these ratios are known to be effective proxies for the ionization parameter as well as being potential signatures of low metallicity gas in galaxies \citep[e.g.][]{nakajima2014,cullen2016,sanders2016,strom2018}.
Similarly, \citet{erb2016} have shown that \lya \ emission is stronger in highly-ionized, low metallicity galaxies selected via their high \oiii/\hbeta \ and low \nii/\halpha \ ratios.
Comparable results have also been found using local `Green Pea' galaxies \citep{yang2017}.
Generally, studies that probe gas-phase metallicity find that \lya \ emission is enhanced in low metallicity environments \citep[e.g.][]{finkelstein2011,nakajima2013,du2019}.
Our results add further support to this picture, by explicitly demonstrating that \wlya \ increases in galaxies with lower stellar metallicity populations and, therefore, harder ionizing radiation fields.

Finally, another important factor in determining \lya \ escape is the dust content of galaxies.
Dust absorbs and scatters \lya \ photons and therefore galaxies with higher dust covering fractions should have lower \wlya.
Indeed, this correlation has been demonstrated in a number of different studies \citep[e.g.][]{kornei2010,pentericci2010,marchi2019,sobral2019}.
Using the global shape of the composite spectra we can roughly estimate the typical FUV dust attenuation in our \wlya \ quartiles.
The FUV continuum slope of a galaxy, $\beta$, (where $f_{\lambda}\propto \lambda^{\beta}$) is known to be an effective proxy for the global dust attenuation at all redshifts, with bluer slopes indicating less dust \citep[e.g.][]{meurer1999,cullen2017}\footnote{Although the intrinsic UV slope also has a dependence on \zstar \ and stellar population age \citep[e.g.][]{castellano2014,rogers2014}, dust attenuation should be the dominant factor in determining the observed $\beta$ value for typical star-forming galaxies at these redshifts \citep[e.g.][]{cullen2017}.}. 
$\beta$ values were measured for each of the composite spectra following the method outlined in \citet{cullen2017} and are given in Table \ref{table:lya_stack_properties}.
The slopes clearly become bluer (i.e. steeper) as \wlya \ increases (as can also be clearly seen in Fig. \ref{fig:fuv_fits}).
Converting these $\beta$ values into dust attenuation at $1500 \rm{\AA}$ following the prescription of \citet{cullen2017} indicates that $\rm{A}_{1500}$ decreases by a factor $\simeq 5$ between the highest and lowest \wlya \ quartiles.

\subsection{The relative importance of intrinsic production versus escape}

 \begin{table}
    \centering
    \caption{The ionizing continuum photon production rate ($N_{\rm{ion}}$) and resulting intrinsic \wlya \ estimated from the best-fitting stellar population models in each \wlya \ quartile.
    Values are calculated for the Starburst99 models used in this paper and also for the BPASSv2.2 models \citep{eldridge2017,stanway2018} assuming the same star formation history and best-fitting stellar metallicity.}\label{table:intrinsic_wlya}
    \begin{tabular}{cccc}
        \hline
        \hline
        \multicolumn{4}{c}{Starburst99}\\
        \hline
        Qartile & \logz & log($N_{\rm{ion}}/\rm{s}^{-1}$) & \wlya$_{\rm{int}}/\rm{\AA}$ \\
        \hline
        Q1 & $-0.69$ & $53.28$  & $102$ \\
        Q2 & $-0.82$ & $53.31$  & $106$ \\
        Q3 & $-0.98$ & $52.32$  & $107$ \\
        Q4 & $<1.08$ & $>52.32$ & $>107$ \\
        \hline
        \hline
        \multicolumn{4}{c}{BPASS v2.2}\\
        \hline
        Q1 & $-0.69$ & $53.58$  & $117$ \\
        Q2 & $-0.82$ & $53.60$  & $121$ \\
        Q3 & $-0.98$ & $53.62$  & $125$ \\
        Q4 & $<1.08$ & $>53.63$ & $>127$ \\
        \hline
    \end{tabular}
    \end{table}

    \begin{figure}
        \centerline{\includegraphics[width=\columnwidth]{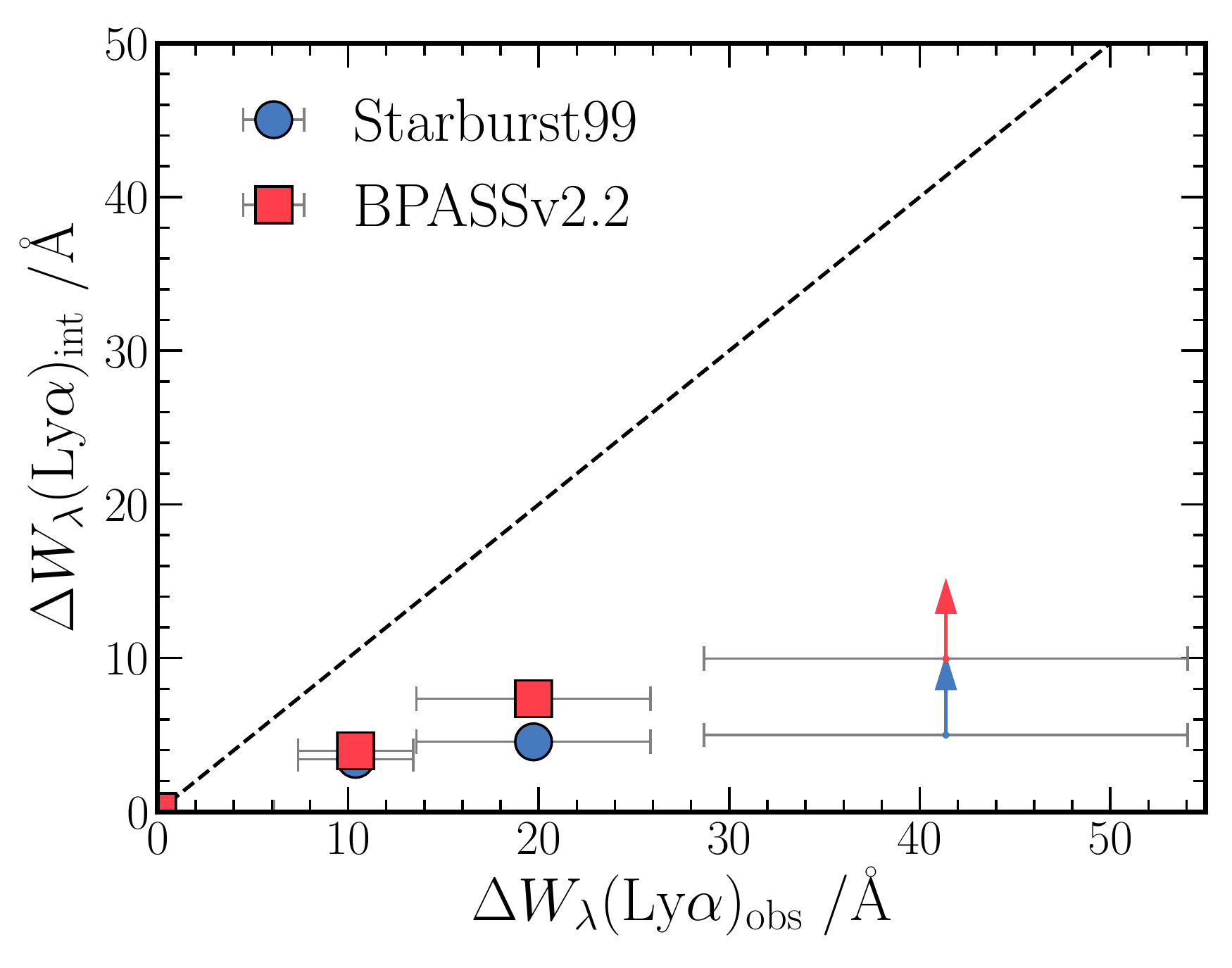}}
        \caption{Intrinsic difference in \lya \ equivalent width \ ($\Delta$\wlya$_{\rm{int}}$) as a function of the observed difference ($\Delta$\wlya$_{\rm{obs}}$) between Q$_{\rm{N}}$ and Q4.
        For the blue circular data points, $\Delta$\wlya$_{\rm{int}}$ was calculated based on the ionizing continuum properties of the best-fitting Starburst99 model (see text for details).
        For the red square data points, $\Delta$\wlya$_{\rm{int}}$ was calculated using the BPASSv2.2 models assuming the same star formation history and best-fitting stellar metallicity.
        If the data fell on the 1:1 relation (dashed black line) this would imply that changes in the ionizing production efficiency of the stellar population with \zstar \ alone could account for the observed variation in \wlya.}
        \label{fig:intrinsic_ew_evolution} 
    \end{figure}

While it is clear that our results are consistent with a picture in which the observed \wlya \ depends both upon the intrinsic production rate of \lya \ photons and on the overall \lya \ opacity (or equivalently the \lya \ escape fraction), we can also attempt to estimate the relative importance of these two physical effects.
For each \wlya \ quartile  we first determined the rate of ionizing photon emission ($N_{\rm{ion}}$ [s$^{-1}$]) from the best-fitting Starburst99 model by integrating the spectrum below $912 \rm{\AA}$.
Then, assuming a simple conversion between $N_{\rm{ion}}$ and \halpha \ luminosity \citep{kennicutt1998}  and an intrinsic \lya/\halpha \ ratio of 8.7 \citep{osterbrock1989}, we estimated the \lya \ luminosity as 
\begin{equation}
L(\mathrm{Ly}\alpha) [\mathrm{erg}\mathrm{s}^{-1}]=1.18\times10^{-11}N_{\rm{ion}}[\mathrm{s}^{-1}].
\end{equation}
The continuum luminosity density ($L_{\lambda,\rm{UV}}$) was defined as the median model luminosity density between $1228-1255\rm{\AA}$ and the intrinsic equivalent width (\wlya$_{\rm{int}}$) estimated as $L(\mathrm{Ly}\alpha)/L_{\lambda,\rm{UV}}$.
Values for $N_{\rm{ion}}$ and \wlya$_{\rm{int}}$ are given in Table \ref{table:intrinsic_wlya}.
We also report, in Table \ref{table:intrinsic_wlya}, the same values calculated using the BPASSv2.2 SPS models \citep{eldridge2017,stanway2018}, where we have assumed the same star formation history and best-fitting metallicity as for the Starburst99 models.
We note that for Q1, since we can only estimate and upper limit on \zstar, we can also only estimate a lower limit on \wlya$_{\rm{int}}$.
We also note that this analysis assumes a $0\%$ escape fraction of ionizing continuum photons ($f_{\rm{esc}}=0$). 
However, given the low average escape fraction of galaxies at these redshifts \citep[e.g. $f_{\rm{esc}}=0.09\pm0.01$,][]{steidel2018}, for the purpose of this discussion it should be a reasonable assumption.

It can clearly be seen that the values of \wlya$_{\rm{int}}$ reported in Table \ref{table:intrinsic_wlya} are much larger than the observed \wlya \ values in Table \ref{table:lya_stack_properties}, which is unsurprising given the relatively large \lya \ opacities expected in general.
Perhaps more interesting is the fact that the differences in \wlya$_{\rm{int}}$ across the quartiles $-$ which are due exclusively to changes in the ionizing continuum strength with \zstar \ $-$ are much smaller than the observed differences in \wlya.
This is clearly illustrated in Fig. \ref{fig:intrinsic_ew_evolution}.
For the Starburst99 models, we estimate that \wlya$_{\rm{int}}$ varies by $\simeq 5\rm{A}$ between Q4 and Q1, which accounts for only $\simeq 12 \%$ of the total observed variation ($\simeq 40 \rm{\AA}$). 
The value is slightly larger assuming the BPASSv2.2 models ($\simeq 10\rm{A}$) but is still a minority effect ($\simeq 25 \%$).

This result suggests that, on average, the change in \wlya \ across the quartiles is being driven primarily by a variation in the \lya \ escape fraction in low \zstar \ galaxies ($\simeq75-85\%$ contribution) as opposed to the intrinsic production rate of \lya \ photons ($\simeq15-25\%$ contribution).
Based on this picture, the strong correlation between \wlya \ and \zstar \ we observe, which results in low \zstar \ galaxies exhibiting stronger \lya \ emission, is a result of three factors: (i) an increase in the production rate of \lya \ photons at lower \zstar, (ii) a decrease in the covering fraction of \hi \ gas due to stronger ionizing continua at lower \zstar \ and, (iii) a decrease in the overall dust content of galaxies at lower \zstar, with the combination of (ii) and (iii) providing the dominant contribution to the observed relation.
Finally, we stress that these conclusions apply to the star-forming population on average, and assume that 100 Myr constant star formation histories are a reasonable approximation for the majority of the FUV spectra at these redshifts \citep{steidel2016,cullen2019}.
For individual galaxies with bursty star-formation histories and UV-ages $\lesssim20$ Myr (e.g. galaxies with the largest \wlya; Fig \ref{fig:simulation}) there may also be age-dependent affects governing the relative $N_{\rm{ion}}$ \citep[e.g.][]{stanway2016}.

\section{Conclusions}\label{sec:conclusion}

In this paper we have presented, for the first time, an investigation into the correlation between \lya \ equivalent width and stellar metallicity for a sample of 768 star-forming galaxies at $3 \leq z \leq 5$ drawn from the VANDELS survey \citep{mclure_vandels,pentericci_vandels}.
Our main results can be summarised as follows:

\begin{enumerate}

    \item Splitting our sample into four \wlya \ quartiles we observe a strong anti-correlation between \wlya \ and \zstar.
    We find that \zstar \ decreases by a factor $\gtrsim 3$ between the lowest \wlya \ quartile ($\langle$\wlya$\rangle=-18\rm{\AA}$) and the highest \wlya \ quartile ($\langle$\wlya$\rangle=24\rm{\AA}$).
    \vspace{2mm}
    \item The same relation is observed if we split our sample into LAEs (\wlya \ $>20\rm{\AA}$) and non-LAEs (\wlya \ $\leq20\rm{\AA}$). 
    On average, the non-LAEs in our sample are $\gtrsim 2 \times$ more metal enriched than the LAE population. 
    \vspace{2mm}
    \item Employing a simple simulation, we show that the \wlya-\logz \ relation presented here, in combination with the stellar MZR presented in \citet{cullen2019}, can reproduce the observed \wlya-\logm \ distribution for $\simeq 95\%$ of our sample. 
    Crucially, however, this simple model fails to account for the $\simeq 5\%$ of our sample with \wlya \ $\gtrsim 50\rm{\AA}$ (and typically with \logm \ $\lesssim9.5$). 
    This result could indicate that our assumption of a constant star-formation history breaks down for some individual galaxies at the lowest stellar masses, where bursty star-formation histories may become more prevalent.
    \vspace{2mm}
    \item We observe a clear correlation between \wlya \ and \wciii \ consistent with previous measurements at similar redshifts. 
    Our results indicate that the strength of both lines increases with decreasing stellar metallicity.
    This provides further evidence to support the idea that the harder ionizing continuum spectra emitted by low metallicity stellar populations plays a role in modulating both the emergent \lya \ and \ciii \ emission in star-forming galaxies.
    \vspace{2mm}
    \item Finally, by estimating the intrinsic \lya \ equivalent widths (\wlya$_{\rm{int}}$) for each quartile, we show that the contribution to the observed variation of \wlya \ due to changes in the ionizing spectrum with \zstar \ is of the order $\simeq 15-25\%$.
    The dominant contribution ($75-85\%$) is therefore a variation in the \lya \ opacity (or escape fraction) with \zstar, presumably due to a combination of lower \hi \ and dust covering fractions in low \zstar \ galaxies.

\end{enumerate}

Overall, the results presented here provide further evidence$-$using, for the first time, direct estimates of \zstar$-$ for a scenario in which low-mass, less dust obscured, galaxies with low-metallicity ionizing stellar populations are both the most efficient producers of \lya \ photons, and the systems from which those photons have the highest likelihood of escape.

\section{Acknowledgments}
FC, RJM, JSD, AC and DJM acknowledge the support of the UK Science and Technology Facilities Council.
A. Cimatti acknowledges the grants ASI n.2018-23-HH.0, PRIN MIUR 2015 and PRIN MIUR 2017 - 20173ML3WW 001.
This work is based on data products from observations made with ESO Telescopes at La Silla Paranal Observatory under ESO programme ID 194.A-2003(E-Q).
We thank the referee for useful suggestions that have improved this paper.
This research made use of Astropy, a community-developed core Python package for Astronomy \citep{astropy2018}, NumPy and SciPy \citep{oliphant2007}, Matplotlib \citep{hunter2007}, {IPython} \citep{perez2007} and NASA's Astrophysics Data System Bibliographic Services. 

\bibliographystyle{mnras}        
\bibliography{vandels_lya}       

\begin{thebibliography}{}
\makeatletter
\relax
\def\mn@urlcharsother{\let\do\@makeother \do\$\do\&\do\#\do\^\do\_\do\%\do\~}
\def\mn@doi{\begingroup\mn@urlcharsother \@ifnextchar [ {\mn@doi@}
  {\mn@doi@[]}}
\def\mn@doi@[#1]#2{\def\@tempa{#1}\ifx\@tempa\@empty \href
  {http://dx.doi.org/#2} {doi:#2}\else \href {http://dx.doi.org/#2} {#1}\fi
  \endgroup}
\def\mn@eprint#1#2{\mn@eprint@#1:#2::\@nil}
\def\mn@eprint@arXiv#1{\href {http://arxiv.org/abs/#1} {{\tt arXiv:#1}}}
\def\mn@eprint@dblp#1{\href {http://dblp.uni-trier.de/rec/bibtex/#1.xml}
  {dblp:#1}}
\def\mn@eprint@#1:#2:#3:#4\@nil{\def\@tempa {#1}\def\@tempb {#2}\def\@tempc
  {#3}\ifx \@tempc \@empty \let \@tempc \@tempb \let \@tempb \@tempa \fi \ifx
  \@tempb \@empty \def\@tempb {arXiv}\fi \@ifundefined
  {mn@eprint@\@tempb}{\@tempb:\@tempc}{\expandafter \expandafter \csname
  mn@eprint@\@tempb\endcsname \expandafter{\@tempc}}}

\bibitem[\protect\citeauthoryear{{Asplund}, {Grevesse}, {Sauval}  \&
  {Scott}}{{Asplund} et~al.}{2009}]{asplund2009}
{Asplund} M.,  {Grevesse} N.,  {Sauval} A.~J.,   {Scott} P.,  2009, \mn@doi
  [Annual Review of Astronomy and Astrophysics]
  {10.1146/annurev.astro.46.060407.145222}, \href
  {https://ui.adsabs.harvard.edu/\#abs/2009ARA&A..47..481A} {47, 481}

\bibitem[\protect\citeauthoryear{{Astropy Collaboration} et~al.,}{{Astropy
  Collaboration} et~al.}{2018}]{astropy2018}
{Astropy Collaboration} et~al., 2018, \mn@doi [\aj] {10.3847/1538-3881/aabc4f},
  \href {https://ui.adsabs.harvard.edu/abs/2018AJ....156..123A} {156, 123}

\bibitem[\protect\citeauthoryear{{Bruzual} \& {Charlot}}{{Bruzual} \&
  {Charlot}}{2003}]{bruzual2003}
{Bruzual} G.,  {Charlot} S.,  2003, \mn@doi [\mnras]
  {10.1046/j.1365-8711.2003.06897.x}, \href
  {http://adsabs.harvard.edu/abs/2003MNRAS.344.1000B} {344, 1000}

\bibitem[\protect\citeauthoryear{{Buchner} et~al.,}{{Buchner}
  et~al.}{2014}]{buchner2014}
{Buchner} J.,  et~al., 2014, \mn@doi [\aap] {10.1051/0004-6361/201322971},
  \href {https://ui.adsabs.harvard.edu/\#abs/2014A&A...564A.125B} {564, A125}

\bibitem[\protect\citeauthoryear{{Calzetti}, {Armus}, {Bohlin}, {Kinney},
  {Koornneef}  \& {Storchi-Bergmann}}{{Calzetti} et~al.}{2000}]{calzetti2000}
{Calzetti} D.,  {Armus} L.,  {Bohlin} R.~C.,  {Kinney} A.~L.,  {Koornneef} J.,
   {Storchi-Bergmann} T.,  2000, \mn@doi [\apj] {10.1086/308692}, \href
  {https://ui.adsabs.harvard.edu/abs/2000ApJ...533..682C} {533, 682}

\bibitem[\protect\citeauthoryear{{Cassata} et~al.,}{{Cassata}
  et~al.}{2015}]{cassata2015}
{Cassata} P.,  et~al., 2015, \mn@doi [\aap] {10.1051/0004-6361/201423824},
  \href {https://ui.adsabs.harvard.edu/abs/2015A&A...573A..24C} {573, A24}

\bibitem[\protect\citeauthoryear{{Castellano} et~al.,}{{Castellano}
  et~al.}{2014}]{castellano2014}
{Castellano} M.,  et~al., 2014, \mn@doi [\aap] {10.1051/0004-6361/201322704},
  \href {https://ui.adsabs.harvard.edu/abs/2014A&A...566A..19C} {566, A19}

\bibitem[\protect\citeauthoryear{{Chabrier}}{{Chabrier}}{2003}]{chabrier2003}
{Chabrier} G.,  2003, \mn@doi [\pasp] {10.1086/376392}, \href
  {https://ui.adsabs.harvard.edu/abs/2003PASP..115..763C} {115, 763}

\bibitem[\protect\citeauthoryear{{Charlot} \& {Fall}}{{Charlot} \&
  {Fall}}{1993}]{charlot1993}
{Charlot} S.,  {Fall} S.~M.,  1993, \mn@doi [\apj] {10.1086/173187}, \href
  {https://ui.adsabs.harvard.edu/abs/1993ApJ...415..580C} {415, 580}

\bibitem[\protect\citeauthoryear{{Cullen}, {Cirasuolo}, {McLure}, {Dunlop}  \&
  {Bowler}}{{Cullen} et~al.}{2014}]{cullen2014}
{Cullen} F.,  {Cirasuolo} M.,  {McLure} R.~J.,  {Dunlop} J.~S.,   {Bowler}
  R.~A.~A.,  2014, \mn@doi [\mnras] {10.1093/mnras/stu443}, \href
  {https://ui.adsabs.harvard.edu/abs/2014MNRAS.440.2300C} {440, 2300}

\bibitem[\protect\citeauthoryear{{Cullen}, {Cirasuolo}, {Kewley}, {McLure},
  {Dunlop}  \& {Bowler}}{{Cullen} et~al.}{2016}]{cullen2016}
{Cullen} F.,  {Cirasuolo} M.,  {Kewley} L.~J.,  {McLure} R.~J.,  {Dunlop}
  J.~S.,   {Bowler} R.~A.~A.,  2016, \mn@doi [\mnras] {10.1093/mnras/stw1181},
  \href {https://ui.adsabs.harvard.edu/abs/2016MNRAS.460.3002C} {460, 3002}

\bibitem[\protect\citeauthoryear{{Cullen}, {McLure}, {Khochfar}, {Dunlop}  \&
  {Dalla Vecchia}}{{Cullen} et~al.}{2017}]{cullen2017}
{Cullen} F.,  {McLure} R.~J.,  {Khochfar} S.,  {Dunlop} J.~S.,   {Dalla
  Vecchia} C.,  2017, \mn@doi [\mnras] {10.1093/mnras/stx1451}, \href
  {http://adsabs.harvard.edu/abs/2017MNRAS.470.3006C} {470, 3006}

\bibitem[\protect\citeauthoryear{{Cullen} et~al.,}{{Cullen}
  et~al.}{2019}]{cullen2019}
{Cullen} F.,  et~al., 2019, \mn@doi [\mnras] {10.1093/mnras/stz1402}, \href
  {https://ui.adsabs.harvard.edu/abs/2019MNRAS.487.2038C} {487, 2038}

\bibitem[\protect\citeauthoryear{{Dijkstra}}{{Dijkstra}}{2014}]{dijkstra2014}
{Dijkstra} M.,  2014, \mn@doi [\pasa] {10.1017/pasa.2014.33}, \href
  {https://ui.adsabs.harvard.edu/abs/2014PASA...31...40D} {31, e040}

\bibitem[\protect\citeauthoryear{{Du} et~al.,}{{Du} et~al.}{2018}]{du2018}
{Du} X.,  et~al., 2018, \mn@doi [\apj] {10.3847/1538-4357/aabfcf}, \href
  {https://ui.adsabs.harvard.edu/abs/2018ApJ...860...75D} {860, 75}

\bibitem[\protect\citeauthoryear{{Du}, {Shapley}, {Tang}, {Stark}, {Martin},
  {Mobasher}, {Topping}  \& {Chevallard}}{{Du} et~al.}{2019}]{du2019}
{Du} X.,  {Shapley} A.~E.,  {Tang} M.,  {Stark} D.~P.,  {Martin} C.~L.,
  {Mobasher} B.,  {Topping} M.~W.,   {Chevallard} J.,  2019, arXiv e-prints,
  \href {https://ui.adsabs.harvard.edu/abs/2019arXiv191011877D} {p.
  arXiv:1910.11877}

\bibitem[\protect\citeauthoryear{{Eldridge}, {Stanway}, {Xiao}, {McClelland },
  {Taylor}, {Ng}, {Greis}  \& {Bray}}{{Eldridge} et~al.}{2017}]{eldridge2017}
{Eldridge} J.~J.,  {Stanway} E.~R.,  {Xiao} L.,  {McClelland } L.~A.~S.,
  {Taylor} G.,  {Ng} M.,  {Greis} S.~M.~L.,   {Bray} J.~C.,  2017, \mn@doi
  [\pasa] {10.1017/pasa.2017.51}, \href
  {https://ui.adsabs.harvard.edu/abs/2017PASA...34...58E} {34, e058}

\bibitem[\protect\citeauthoryear{{Erb}, {Pettini}, {Shapley}, {Steidel}, {Law}
  \& {Reddy}}{{Erb} et~al.}{2010}]{erb2010}
{Erb} D.~K.,  {Pettini} M.,  {Shapley} A.~E.,  {Steidel} C.~C.,  {Law} D.~R.,
  {Reddy} N.~A.,  2010, \mn@doi [\apj] {10.1088/0004-637X/719/2/1168}, \href
  {https://ui.adsabs.harvard.edu/abs/2010ApJ...719.1168E} {719, 1168}

\bibitem[\protect\citeauthoryear{{Erb} et~al.,}{{Erb} et~al.}{2014}]{erb2014}
{Erb} D.~K.,  et~al., 2014, \mn@doi [\apj] {10.1088/0004-637X/795/1/33}, \href
  {https://ui.adsabs.harvard.edu/abs/2014ApJ...795...33E} {795, 33}

\bibitem[\protect\citeauthoryear{{Erb}, {Pettini}, {Steidel}, {Strom}, {Rudie},
  {Trainor}, {Shapley}  \& {Reddy}}{{Erb} et~al.}{2016}]{erb2016}
{Erb} D.~K.,  {Pettini} M.,  {Steidel} C.~C.,  {Strom} A.~L.,  {Rudie} G.~C.,
  {Trainor} R.~F.,  {Shapley} A.~E.,   {Reddy} N.~A.,  2016, \mn@doi [\apj]
  {10.3847/0004-637X/830/1/52}, \href
  {https://ui.adsabs.harvard.edu/abs/2016ApJ...830...52E} {830, 52}

\bibitem[\protect\citeauthoryear{{Feroz} \& {Hobson}}{{Feroz} \&
  {Hobson}}{2008}]{feroz2008}
{Feroz} F.,  {Hobson} M.~P.,  2008, \mn@doi [\mnras]
  {10.1111/j.1365-2966.2007.12353.x}, \href
  {https://ui.adsabs.harvard.edu/\#abs/2008MNRAS.384..449F} {384, 449}

\bibitem[\protect\citeauthoryear{{Feroz}, {Hobson}  \& {Bridges}}{{Feroz}
  et~al.}{2009}]{feroz2009}
{Feroz} F.,  {Hobson} M.~P.,   {Bridges} M.,  2009, \mn@doi [\mnras]
  {10.1111/j.1365-2966.2009.14548.x}, \href
  {https://ui.adsabs.harvard.edu/\#abs/2009MNRAS.398.1601F} {398, 1601}

\bibitem[\protect\citeauthoryear{{Finkelstein} et~al.,}{{Finkelstein}
  et~al.}{2011}]{finkelstein2011}
{Finkelstein} S.~L.,  et~al., 2011, \mn@doi [\apj]
  {10.1088/0004-637X/729/2/140}, \href
  {https://ui.adsabs.harvard.edu/abs/2011ApJ...729..140F} {729, 140}

\bibitem[\protect\citeauthoryear{{Fontanot}, {Cristiani}, {Pfrommer}, {Cupani}
  \& {Vanzella}}{{Fontanot} et~al.}{2014}]{fontanot2014}
{Fontanot} F.,  {Cristiani} S.,  {Pfrommer} C.,  {Cupani} G.,   {Vanzella} E.,
  2014, \mn@doi [\mnras] {10.1093/mnras/stt2332}, \href
  {https://ui.adsabs.harvard.edu/abs/2014MNRAS.438.2097F} {438, 2097}

\bibitem[\protect\citeauthoryear{{Gallazzi}, {Charlot}, {Brinchmann}, {White}
  \& {Tremonti}}{{Gallazzi} et~al.}{2005}]{gallazzi2005}
{Gallazzi} A.,  {Charlot} S.,  {Brinchmann} J.,  {White} S. D.~M.,   {Tremonti}
  C.~A.,  2005, \mn@doi [\mnras] {10.1111/j.1365-2966.2005.09321.x}, \href
  {https://ui.adsabs.harvard.edu/abs/2005MNRAS.362...41G} {362, 41}

\bibitem[\protect\citeauthoryear{{Grogin} et~al.,}{{Grogin}
  et~al.}{2011}]{grogin2011}
{Grogin} N.~A.,  et~al., 2011, \mn@doi [\apjs] {10.1088/0067-0049/197/2/35},
  \href {https://ui.adsabs.harvard.edu/abs/2011ApJS..197...35G} {197, 35}

\bibitem[\protect\citeauthoryear{{Hathi} et~al.,}{{Hathi}
  et~al.}{2016}]{hathi2016}
{Hathi} N.~P.,  et~al., 2016, \mn@doi [\aap] {10.1051/0004-6361/201526012},
  \href {https://ui.adsabs.harvard.edu/abs/2016A&A...588A..26H} {588, A26}

\bibitem[\protect\citeauthoryear{{Hayes} et~al.,}{{Hayes}
  et~al.}{2014}]{hayes2014}
{Hayes} M.,  et~al., 2014, \mn@doi [\apj] {10.1088/0004-637X/782/1/6}, \href
  {https://ui.adsabs.harvard.edu/abs/2014ApJ...782....6H} {782, 6}

\bibitem[\protect\citeauthoryear{{Heckman} et~al.,}{{Heckman}
  et~al.}{2011}]{heckman2011}
{Heckman} T.~M.,  et~al., 2011, \mn@doi [\apj] {10.1088/0004-637X/730/1/5},
  \href {https://ui.adsabs.harvard.edu/abs/2011ApJ...730....5H} {730, 5}

\bibitem[\protect\citeauthoryear{Hunter}{Hunter}{2007}]{hunter2007}
Hunter J.~D.,  2007, Computing In Science \& Engineering, 9, 90

\bibitem[\protect\citeauthoryear{{Jaskot} \& {Ravindranath}}{{Jaskot} \&
  {Ravindranath}}{2016}]{jaskot2016}
{Jaskot} A.~E.,  {Ravindranath} S.,  2016, \mn@doi [\apj]
  {10.3847/1538-4357/833/2/136}, \href
  {https://ui.adsabs.harvard.edu/abs/2016ApJ...833..136J} {833, 136}

\bibitem[\protect\citeauthoryear{{Kennicutt}}{{Kennicutt}}{1998}]{kennicutt1998}
{Kennicutt} Robert~C. J.,  1998, \mn@doi [\araa]
  {10.1146/annurev.astro.36.1.189}, \href
  {https://ui.adsabs.harvard.edu/abs/1998ARA&A..36..189K} {36, 189}

\bibitem[\protect\citeauthoryear{{Kewley}, {Dopita}, {Leitherer}, {Dav{\'e}},
  {Yuan}, {Allen}, {Groves}  \& {Sutherland}}{{Kewley}
  et~al.}{2013}]{kewley2013}
{Kewley} L.~J.,  {Dopita} M.~A.,  {Leitherer} C.,  {Dav{\'e}} R.,  {Yuan} T.,
  {Allen} M.,  {Groves} B.,   {Sutherland} R.,  2013, \mn@doi [\apj]
  {10.1088/0004-637X/774/2/100}, \href
  {https://ui.adsabs.harvard.edu/abs/2013ApJ...774..100K} {774, 100}

\bibitem[\protect\citeauthoryear{{Kewley}, {Nicholls}  \&
  {Sutherland}}{{Kewley} et~al.}{2019}]{kewley2019}
{Kewley} L.~J.,  {Nicholls} D.~C.,   {Sutherland} R.~S.,  2019, \mn@doi [\araa]
  {10.1146/annurev-astro-081817-051832}, \href
  {https://ui.adsabs.harvard.edu/abs/2019ARA&A..57..511K} {57, 511}

\bibitem[\protect\citeauthoryear{{Koekemoer} et~al.,}{{Koekemoer}
  et~al.}{2011}]{koekemoer2011}
{Koekemoer} A.~M.,  et~al., 2011, \mn@doi [\apjs] {10.1088/0067-0049/197/2/36},
  \href {https://ui.adsabs.harvard.edu/abs/2011ApJS..197...36K} {197, 36}

\bibitem[\protect\citeauthoryear{{Kornei}, {Shapley}, {Erb}, {Steidel},
  {Reddy}, {Pettini}  \& {Bogosavljevi{\'c}}}{{Kornei}
  et~al.}{2010}]{kornei2010}
{Kornei} K.~A.,  {Shapley} A.~E.,  {Erb} D.~K.,  {Steidel} C.~C.,  {Reddy}
  N.~A.,  {Pettini} M.,   {Bogosavljevi{\'c}} M.,  2010, \mn@doi [\apj]
  {10.1088/0004-637X/711/2/693}, \href
  {https://ui.adsabs.harvard.edu/abs/2010ApJ...711..693K} {711, 693}

\bibitem[\protect\citeauthoryear{{Kriek}, {van Dokkum}, {Labb{\'e}}, {Franx},
  {Illingworth}, {Marchesini}  \& {Quadri}}{{Kriek} et~al.}{2009}]{kriek2009}
{Kriek} M.,  {van Dokkum} P.~G.,  {Labb{\'e}} I.,  {Franx} M.,  {Illingworth}
  G.~D.,  {Marchesini} D.,   {Quadri} R.~F.,  2009, \mn@doi [\apj]
  {10.1088/0004-637X/700/1/221}, \href
  {https://ui.adsabs.harvard.edu/\#abs/2009ApJ...700..221K} {700, 221}

\bibitem[\protect\citeauthoryear{{Law}, {Steidel}, {Shapley}, {Nagy}, {Reddy}
  \& {Erb}}{{Law} et~al.}{2012}]{law2012}
{Law} D.~R.,  {Steidel} C.~C.,  {Shapley} A.~E.,  {Nagy} S.~R.,  {Reddy} N.~A.,
    {Erb} D.~K.,  2012, \mn@doi [\apj] {10.1088/0004-637X/759/1/29}, \href
  {https://ui.adsabs.harvard.edu/abs/2012ApJ...759...29L} {759, 29}

\bibitem[\protect\citeauthoryear{{Le F{\`e}vre} et~al.,}{{Le F{\`e}vre}
  et~al.}{2019}]{lefevre2019}
{Le F{\`e}vre} O.,  et~al., 2019, \mn@doi [\aap] {10.1051/0004-6361/201732197},
  \href {https://ui.adsabs.harvard.edu/abs/2019A&A...625A..51L} {625, A51}

\bibitem[\protect\citeauthoryear{{Leitherer}, {Ortiz Ot{\'a}lvaro}, {Bresolin},
  {Kudritzki}, {Lo Faro}, {Pauldrach}, {Pettini}  \& {Rix}}{{Leitherer}
  et~al.}{2010}]{leitherer2010}
{Leitherer} C.,  {Ortiz Ot{\'a}lvaro} P.~A.,  {Bresolin} F.,  {Kudritzki}
  R.-P.,  {Lo Faro} B.,  {Pauldrach} A. W.~A.,  {Pettini} M.,   {Rix} S.~A.,
  2010, \mn@doi [\apjs] {10.1088/0067-0049/189/2/309}, \href
  {https://ui.adsabs.harvard.edu/abs/2010ApJS..189..309L} {189, 309}

\bibitem[\protect\citeauthoryear{{Marchi} et~al.,}{{Marchi}
  et~al.}{2018}]{marchi2018}
{Marchi} F.,  et~al., 2018, \mn@doi [\aap] {10.1051/0004-6361/201732133}, \href
  {https://ui.adsabs.harvard.edu/abs/2018A&A...614A..11M} {614, A11}

\bibitem[\protect\citeauthoryear{{Marchi} et~al.,}{{Marchi}
  et~al.}{2019}]{marchi2019}
{Marchi} F.,  et~al., 2019, \mn@doi [\aap] {10.1051/0004-6361/201935495}, \href
  {https://ui.adsabs.harvard.edu/abs/2019A&A...631A..19M} {631, A19}

\bibitem[\protect\citeauthoryear{{Matthee}, {Sobral}, {Darvish}, {Santos},
  {Mobasher}, {Paulino-Afonso}, {R{\"o}ttgering}  \& {Alegre}}{{Matthee}
  et~al.}{2017}]{matthee2017}
{Matthee} J.,  {Sobral} D.,  {Darvish} B.,  {Santos} S.,  {Mobasher} B.,
  {Paulino-Afonso} A.,  {R{\"o}ttgering} H.,   {Alegre} L.,  2017, \mn@doi
  [\mnras] {10.1093/mnras/stx2061}, \href
  {https://ui.adsabs.harvard.edu/abs/2017MNRAS.472..772M} {472, 772}

\bibitem[\protect\citeauthoryear{{McLure} et~al.,}{{McLure}
  et~al.}{2018}]{mclure_vandels}
{McLure} R.~J.,  et~al., 2018, \mn@doi [\mnras] {10.1093/mnras/sty1213}, \href
  {https://ui.adsabs.harvard.edu/abs/2018MNRAS.479...25M} {479, 25}

\bibitem[\protect\citeauthoryear{{Meurer}, {Heckman}  \& {Calzetti}}{{Meurer}
  et~al.}{1999}]{meurer1999}
{Meurer} G.~R.,  {Heckman} T.~M.,   {Calzetti} D.,  1999, \mn@doi [\apj]
  {10.1086/307523}, \href {http://adsabs.harvard.edu/abs/1999ApJ...521...64M}
  {521, 64}

\bibitem[\protect\citeauthoryear{{Nakajima} \& {Ouchi}}{{Nakajima} \&
  {Ouchi}}{2014}]{nakajima2014}
{Nakajima} K.,  {Ouchi} M.,  2014, \mn@doi [\mnras] {10.1093/mnras/stu902},
  \href {https://ui.adsabs.harvard.edu/abs/2014MNRAS.442..900N} {442, 900}

\bibitem[\protect\citeauthoryear{{Nakajima}, {Ouchi}, {Shimasaku}, {Hashimoto},
  {Ono}  \& {Lee}}{{Nakajima} et~al.}{2013}]{nakajima2013}
{Nakajima} K.,  {Ouchi} M.,  {Shimasaku} K.,  {Hashimoto} T.,  {Ono} Y.,
  {Lee} J.~C.,  2013, \mn@doi [\apj] {10.1088/0004-637X/769/1/3}, \href
  {https://ui.adsabs.harvard.edu/abs/2013ApJ...769....3N} {769, 3}

\bibitem[\protect\citeauthoryear{{Nakajima} et~al.,}{{Nakajima}
  et~al.}{2018}]{nakajima2018}
{Nakajima} K.,  et~al., 2018, \mn@doi [\aap] {10.1051/0004-6361/201731935},
  \href {https://ui.adsabs.harvard.edu/abs/2018A&A...612A..94N} {612, A94}

\bibitem[\protect\citeauthoryear{{Nestor}, {Shapley}, {Kornei}, {Steidel}  \&
  {Siana}}{{Nestor} et~al.}{2013}]{nestor2013}
{Nestor} D.~B.,  {Shapley} A.~E.,  {Kornei} K.~A.,  {Steidel} C.~C.,   {Siana}
  B.,  2013, \mn@doi [\apj] {10.1088/0004-637X/765/1/47}, \href
  {https://ui.adsabs.harvard.edu/abs/2013ApJ...765...47N} {765, 47}

\bibitem[\protect\citeauthoryear{{Noll}, {Burgarella}, {Giovannoli}, {Buat},
  {Marcillac}  \& {Mu{\~n}oz-Mateos}}{{Noll} et~al.}{2009}]{noll2009}
{Noll} S.,  {Burgarella} D.,  {Giovannoli} E.,  {Buat} V.,  {Marcillac} D.,
  {Mu{\~n}oz-Mateos} J.~C.,  2009, \mn@doi [\aap]
  {10.1051/0004-6361/200912497}, \href
  {https://ui.adsabs.harvard.edu/abs/2009A&A...507.1793N} {507, 1793}

\bibitem[\protect\citeauthoryear{Oliphant}{Oliphant}{2007}]{oliphant2007}
Oliphant T.~E.,  2007, Computing in Science \& Engineering, 9, 10

\bibitem[\protect\citeauthoryear{{Osterbrock}}{{Osterbrock}}{1989}]{osterbrock1989}
{Osterbrock} D.~E.,  1989, {Astrophysics of gaseous nebulae and active galactic
  nuclei}

\bibitem[\protect\citeauthoryear{{Oyarz{\'u}n} et~al.,}{{Oyarz{\'u}n}
  et~al.}{2016}]{oyarzum2016}
{Oyarz{\'u}n} G.~A.,  et~al., 2016, \mn@doi [\apjl]
  {10.3847/2041-8205/821/1/L14}, \href
  {https://ui.adsabs.harvard.edu/abs/2016ApJ...821L..14O} {821, L14}

\bibitem[\protect\citeauthoryear{{Oyarz{\'u}n}, {Blanc}, {Gonz{\'a}lez},
  {Mateo}  \& {Bailey}}{{Oyarz{\'u}n} et~al.}{2017}]{oyarzum2017}
{Oyarz{\'u}n} G.~A.,  {Blanc} G.~A.,  {Gonz{\'a}lez} V.,  {Mateo} M.,
  {Bailey} John~I. I.,  2017, \mn@doi [\apj] {10.3847/1538-4357/aa7552}, \href
  {https://ui.adsabs.harvard.edu/abs/2017ApJ...843..133O} {843, 133}

\bibitem[\protect\citeauthoryear{{Pahl}, {Shapley}, {Faisst}, {Capak}, {Du},
  {Reddy}, {Laursen}  \& {Topping}}{{Pahl} et~al.}{2020}]{pahl2020}
{Pahl} A.~J.,  {Shapley} A.,  {Faisst} A.~L.,  {Capak} P.~L.,  {Du} X.,
  {Reddy} N.~A.,  {Laursen} P.,   {Topping} M.~W.,  2020, \mn@doi [\mnras]
  {10.1093/mnras/staa355}, \href
  {https://ui.adsabs.harvard.edu/abs/2020MNRAS.tmp..342P} {}

\bibitem[\protect\citeauthoryear{{Pentericci}, {Grazian}, {Scarlata},
  {Fontana}, {Castellano}, {Giallongo}  \& {Vanzella}}{{Pentericci}
  et~al.}{2010}]{pentericci2010}
{Pentericci} L.,  {Grazian} A.,  {Scarlata} C.,  {Fontana} A.,  {Castellano}
  M.,  {Giallongo} E.,   {Vanzella} E.,  2010, \mn@doi [\aap]
  {10.1051/0004-6361/200913425}, \href
  {https://ui.adsabs.harvard.edu/abs/2010A&A...514A..64P} {514, A64}

\bibitem[\protect\citeauthoryear{{Pentericci} et~al.,}{{Pentericci}
  et~al.}{2018}]{pentericci_vandels}
{Pentericci} L.,  et~al., 2018, \mn@doi [\aap] {10.1051/0004-6361/201833047},
  \href {https://ui.adsabs.harvard.edu/abs/2018A&A...616A.174P} {616, A174}

\bibitem[\protect\citeauthoryear{P\'{e}rez \& Granger}{P\'{e}rez \&
  Granger}{2007}]{perez2007}
P\'{e}rez F.,  Granger B.~E.,  2007, Computing in Science {\&} Engineering, 9,
  21

\bibitem[\protect\citeauthoryear{{Rigby}, {Bayliss}, {Gladders}, {Sharon},
  {Wuyts}, {Dahle}, {Johnson}  \& {Pe{\~n}a-Guerrero}}{{Rigby}
  et~al.}{2015}]{rigby2015}
{Rigby} J.~R.,  {Bayliss} M.~B.,  {Gladders} M.~D.,  {Sharon} K.,  {Wuyts} E.,
  {Dahle} H.,  {Johnson} T.,   {Pe{\~n}a-Guerrero} M.,  2015, \mn@doi [\apjl]
  {10.1088/2041-8205/814/1/L6}, \href
  {https://ui.adsabs.harvard.edu/abs/2015ApJ...814L...6R} {814, L6}

\bibitem[\protect\citeauthoryear{{Robertson}, {Ellis}, {Furlanetto}  \&
  {Dunlop}}{{Robertson} et~al.}{2015}]{robertson2015}
{Robertson} B.~E.,  {Ellis} R.~S.,  {Furlanetto} S.~R.,   {Dunlop} J.~S.,
  2015, \mn@doi [\apjl] {10.1088/2041-8205/802/2/L19}, \href
  {https://ui.adsabs.harvard.edu/abs/2015ApJ...802L..19R} {802, L19}

\bibitem[\protect\citeauthoryear{{Rogers} et~al.,}{{Rogers}
  et~al.}{2014}]{rogers2014}
{Rogers} A.~B.,  et~al., 2014, \mn@doi [\mnras] {10.1093/mnras/stu558}, \href
  {https://ui.adsabs.harvard.edu/abs/2014MNRAS.440.3714R} {440, 3714}

\bibitem[\protect\citeauthoryear{{Salim}, {Boquien}  \& {Lee}}{{Salim}
  et~al.}{2018}]{salim2018}
{Salim} S.,  {Boquien} M.,   {Lee} J.~C.,  2018, \mn@doi [\apj]
  {10.3847/1538-4357/aabf3c}, \href
  {https://ui.adsabs.harvard.edu/\#abs/2018ApJ...859...11S} {859, 11}

\bibitem[\protect\citeauthoryear{{Sanders} et~al.,}{{Sanders}
  et~al.}{2016}]{sanders2016}
{Sanders} R.~L.,  et~al., 2016, \mn@doi [\apj] {10.3847/0004-637X/816/1/23},
  \href {https://ui.adsabs.harvard.edu/abs/2016ApJ...816...23S} {816, 23}

\bibitem[\protect\citeauthoryear{{Schaerer}}{{Schaerer}}{2003}]{schaerer2003}
{Schaerer} D.,  2003, \mn@doi [\aap] {10.1051/0004-6361:20021525}, \href
  {https://ui.adsabs.harvard.edu/abs/2003A&A...397..527S} {397, 527}

\bibitem[\protect\citeauthoryear{{Schaerer}, {Izotov}, {Nakajima}, {Worseck},
  {Chisholm}, {Verhamme}, {Thuan}  \& {de Barros}}{{Schaerer}
  et~al.}{2018}]{schaerer2018}
{Schaerer} D.,  {Izotov} Y.~I.,  {Nakajima} K.,  {Worseck} G.,  {Chisholm} J.,
  {Verhamme} A.,  {Thuan} T.~X.,   {de Barros} S.,  2018, \mn@doi [\aap]
  {10.1051/0004-6361/201833823}, \href
  {https://ui.adsabs.harvard.edu/abs/2018A&A...616L..14S} {616, L14}

\bibitem[\protect\citeauthoryear{{Schreiber} et~al.,}{{Schreiber}
  et~al.}{2018}]{schreiber2018}
{Schreiber} C.,  et~al., 2018, \mn@doi [\aap] {10.1051/0004-6361/201731917},
  \href {https://ui.adsabs.harvard.edu/\#abs/2018A&A...611A..22S} {611, A22}

\bibitem[\protect\citeauthoryear{{Senchyna} et~al.,}{{Senchyna}
  et~al.}{2017}]{senchyna2017}
{Senchyna} P.,  et~al., 2017, \mn@doi [\mnras] {10.1093/mnras/stx2059}, \href
  {https://ui.adsabs.harvard.edu/abs/2017MNRAS.472.2608S} {472, 2608}

\bibitem[\protect\citeauthoryear{{Shapley}, {Steidel}, {Pettini}  \&
  {Adelberger}}{{Shapley} et~al.}{2003}]{shapley2003}
{Shapley} A.~E.,  {Steidel} C.~C.,  {Pettini} M.,   {Adelberger} K.~L.,  2003,
  \mn@doi [\apj] {10.1086/373922}, \href
  {https://ui.adsabs.harvard.edu/abs/2003ApJ...588...65S} {588, 65}

\bibitem[\protect\citeauthoryear{{Sobral} \& {Matthee}}{{Sobral} \&
  {Matthee}}{2019}]{sobral2019}
{Sobral} D.,  {Matthee} J.,  2019, \mn@doi [\aap]
  {10.1051/0004-6361/201833075}, \href
  {https://ui.adsabs.harvard.edu/abs/2019A&A...623A.157S} {623, A157}

\bibitem[\protect\citeauthoryear{{Song} et~al.,}{{Song}
  et~al.}{2014}]{song2014}
{Song} M.,  et~al., 2014, \mn@doi [\apj] {10.1088/0004-637X/791/1/3}, \href
  {https://ui.adsabs.harvard.edu/abs/2014ApJ...791....3S} {791, 3}

\bibitem[\protect\citeauthoryear{{Songaila}}{{Songaila}}{2004}]{songaila2004}
{Songaila} A.,  2004, \mn@doi [\aj] {10.1086/383561}, \href
  {https://ui.adsabs.harvard.edu/abs/2004AJ....127.2598S} {127, 2598}

\bibitem[\protect\citeauthoryear{{Stanway} \& {Eldridge}}{{Stanway} \&
  {Eldridge}}{2018}]{stanway2018}
{Stanway} E.~R.,  {Eldridge} J.~J.,  2018, \mn@doi [\mnras]
  {10.1093/mnras/sty1353}, \href
  {https://ui.adsabs.harvard.edu/abs/2018MNRAS.479...75S} {479, 75}

\bibitem[\protect\citeauthoryear{{Stanway}, {Eldridge}  \& {Becker}}{{Stanway}
  et~al.}{2016}]{stanway2016}
{Stanway} E.~R.,  {Eldridge} J.~J.,   {Becker} G.~D.,  2016, \mn@doi [\mnras]
  {10.1093/mnras/stv2661}, \href
  {https://ui.adsabs.harvard.edu/abs/2016MNRAS.456..485S} {456, 485}

\bibitem[\protect\citeauthoryear{{Stark} et~al.,}{{Stark}
  et~al.}{2014}]{stark2014}
{Stark} D.~P.,  et~al., 2014, \mn@doi [\mnras] {10.1093/mnras/stu1618}, \href
  {https://ui.adsabs.harvard.edu/abs/2014MNRAS.445.3200S} {445, 3200}

\bibitem[\protect\citeauthoryear{{Stark} et~al.,}{{Stark}
  et~al.}{2017}]{stark2017}
{Stark} D.~P.,  et~al., 2017, \mn@doi [\mnras] {10.1093/mnras/stw2233}, \href
  {https://ui.adsabs.harvard.edu/abs/2017MNRAS.464..469S} {464, 469}

\bibitem[\protect\citeauthoryear{Steidel et~al.,}{Steidel
  et~al.}{2014}]{steidel2014}
Steidel C.~C.,  et~al., 2014, \mn@doi [\apj] {10.1088/0004-637X/795/2/165},
  \href {http://adsabs.harvard.edu/abs/2014ApJ...795..165S} {795, 165}

\bibitem[\protect\citeauthoryear{{Steidel}, {Strom}, {Pettini}, {Rudie},
  {Reddy}  \& {Trainor}}{{Steidel} et~al.}{2016}]{steidel2016}
{Steidel} C.~C.,  {Strom} A.~L.,  {Pettini} M.,  {Rudie} G.~C.,  {Reddy} N.~A.,
    {Trainor} R.~F.,  2016, \mn@doi [\apj] {10.3847/0004-637X/826/2/159}, \href
  {http://adsabs.harvard.edu/abs/2016ApJ...826..159S} {826, 159}

\bibitem[\protect\citeauthoryear{{Steidel}, {Bogosavljevi{\'c}}, {Shapley},
  {Reddy}, {Rudie}, {Pettini}, {Trainor}  \& {Strom}}{{Steidel}
  et~al.}{2018}]{steidel2018}
{Steidel} C.~C.,  {Bogosavljevi{\'c}} M.,  {Shapley} A.~E.,  {Reddy} N.~A.,
  {Rudie} G.~C.,  {Pettini} M.,  {Trainor} R.~F.,   {Strom} A.~L.,  2018,
  \mn@doi [\apj] {10.3847/1538-4357/aaed28}, \href
  {https://ui.adsabs.harvard.edu/abs/2018ApJ...869..123S} {869, 123}

\bibitem[\protect\citeauthoryear{{Strom}, {Steidel}, {Rudie}, {Trainor}  \&
  {Pettini}}{{Strom} et~al.}{2018}]{strom2018}
{Strom} A.~L.,  {Steidel} C.~C.,  {Rudie} G.~C.,  {Trainor} R.~F.,   {Pettini}
  M.,  2018, \mn@doi [\apj] {10.3847/1538-4357/aae1a5}, \href
  {https://ui.adsabs.harvard.edu/abs/2018ApJ...868..117S} {868, 117}

\bibitem[\protect\citeauthoryear{{Trainor}, {Steidel}, {Strom}  \&
  {Rudie}}{{Trainor} et~al.}{2015}]{trainor2015}
{Trainor} R.~F.,  {Steidel} C.~C.,  {Strom} A.~L.,   {Rudie} G.~C.,  2015,
  \mn@doi [\apj] {10.1088/0004-637X/809/1/89}, \href
  {https://ui.adsabs.harvard.edu/abs/2015ApJ...809...89T} {809, 89}

\bibitem[\protect\citeauthoryear{{Trainor}, {Strom}, {Steidel}  \&
  {Rudie}}{{Trainor} et~al.}{2016}]{trainor2016}
{Trainor} R.~F.,  {Strom} A.~L.,  {Steidel} C.~C.,   {Rudie} G.~C.,  2016,
  \mn@doi [\apj] {10.3847/0004-637X/832/2/171}, \href
  {https://ui.adsabs.harvard.edu/abs/2016ApJ...832..171T} {832, 171}

\bibitem[\protect\citeauthoryear{{Trainor}, {Strom}, {Steidel}, {Rudie}, {Chen}
   \& {Theios}}{{Trainor} et~al.}{2019}]{trainor2019}
{Trainor} R.~F.,  {Strom} A.~L.,  {Steidel} C.~C.,  {Rudie} G.~C.,  {Chen} Y.,
   {Theios} R.~L.,  2019, \mn@doi [\apj] {10.3847/1538-4357/ab4993}, \href
  {https://ui.adsabs.harvard.edu/abs/2019ApJ...887...85T} {887, 85}

\bibitem[\protect\citeauthoryear{{Verhamme}, {Orlitov{\'a}}, {Schaerer},
  {Izotov}, {Worseck}, {Thuan}  \& {Guseva}}{{Verhamme}
  et~al.}{2017}]{verhamme2017}
{Verhamme} A.,  {Orlitov{\'a}} I.,  {Schaerer} D.,  {Izotov} Y.,  {Worseck} G.,
   {Thuan} T.~X.,   {Guseva} N.,  2017, \mn@doi [\aap]
  {10.1051/0004-6361/201629264}, \href
  {https://ui.adsabs.harvard.edu/abs/2017A&A...597A..13V} {597, A13}

\bibitem[\protect\citeauthoryear{{Yang} et~al.,}{{Yang}
  et~al.}{2017}]{yang2017}
{Yang} H.,  et~al., 2017, \mn@doi [\apj] {10.3847/1538-4357/aa7d4d}, \href
  {https://ui.adsabs.harvard.edu/abs/2017ApJ...844..171Y} {844, 171}

\makeatother
\end{thebibliography}

\appendix

\bigskip

\noindent
$^{1}$SUPA\thanks{Scottish Universities Physics Alliance}, Institute for Astronomy, University of Edinburgh, Royal Observatory, Edinburgh EH9 3HJ\\
$^{2}$Department of Physics and Astronomy, University of California, Los Angeles, 430 Portola Plaza, Los Angeles, CA 90095, USA\\
$^{3}$Instituto de Investigaci\'on Multidisciplinar en Ciencia y Tecnolog\'ia, Universidad de La Serena, Ra\'ul Bitr\'an 1305, La Serena, Chile\\
$^{4}$Departamento de F\'isica y Astronom\'ia, Universidad de La Serena, Av. Juan Cisternas 1200 Norte, La Serena, Chile\\
$^{5}$INAF - Osservatorio Astronomico di Bologna, via P. Gobetti 93/3,I-40129, Bologna, Italy\\
$^{6}$INAF$-$Osservatorio Astronomico di Roma, Via Frascati 33, I-00040 Monte Porzio Catone (RM), Italy\\
$^{7}$University of Bologna, Department of Physics and Astronomy (DIFA) 
Via Gobetti 93/2- 40129, Bologna, Italy\\
$^{8}$INAF - Osservatorio Astrofisico di Arcetri, Largo E. Fermi 5, I-50125, Firenze, Italy\\
$^{9}$European Southern Observatory, Karl-Schwarzschild-Str. 2, 86748 Garching b. M\"unchen, Germany\\
$^{10}$INAF-Astronomical Observatory of Trieste, via G.B. Tiepolo 11, 34143 Trieste, Italy\\
$^{11}$INAF-IASF Milano, via Bassini 15, I-20133, Milano, Italy\\
$^{12}$N\'ucleo de Astronom\'ia, Facultad de Ingenier\'ia, Universidad Diego Portales, Av. Ej\'ercito 441, Santiago, Chile\\
$^{13}$Department of Physics, University of Oxford, Keble Road, Oxford OX1 3RH, UK\\
$^{14}$Department of Physics and Astronomy, University of the Western Cape, Private Bag X17, Bellville, Cape Town, 7535, South Africa
\\
$^{15}$The Cosmic Dawn Center, Niels Bohr Institute, University of Copenhagen, Juliane Maries Vej 30, DK-2100 Copenhagen {{\O}}, Denmark\\
$^{16}$Space Telescope Science Institute, 3700 San Martin Drive, Baltimore, MD 21218, USA\\
$^{17}$Astronomy Department, University of Massachusetts, Amherst, MA 01003, USA

\label{lastpage}
\bsp
\end{document}